\documentclass[sigconf, screen]{acmart}

\AtBeginDocument{%
  }

\setcopyright{none} 
\copyrightyear{}
\acmYear{}
\acmDOI{}

\acmConference[]{}{}{}

\renewcommand\footnotetextcopyrightpermission[1]{}

\usepackage{duckuments}
\usepackage{listings}
\usepackage{xcolor}
\usepackage{float}
\usepackage{caption}
\usepackage{tikz}
\usepackage{xspace}
\usepackage{subcaption}
\usepackage{enumitem}
\usepackage{lipsum}

\definecolor{algobg}{RGB}{246,244,238}      
\definecolor{rulegray}{RGB}{120,120,120}

\definecolor{pragmaBlue}{RGB}{0,70,200}
\definecolor{pragmaRed}{RGB}{200,0,0}
\definecolor{pragmaGreen}{RGB}{0,140,90}
\definecolor{pragmaMagenta}{RGB}{200,0,130}

\definecolor{kwGreen}{RGB}{0,140,90}
\definecolor{kwMagenta}{RGB}{200,0,130}

\lstdefinestyle{hlsalgo}{%
  language=C++,
  basicstyle=\ttfamily\footnotesize,
  columns=fullflexible,
  keepspaces=true,
  showstringspaces=false,
  backgroundcolor=\color{algobg},
  xleftmargin=0em,
  xrightmargin=0em,
  aboveskip=0.6em,
  belowskip=0.2em,
  frame=tb,
  framexleftmargin=10pt,
  rulecolor=\color{black},
  framerule=0.5pt,
  framesep=6pt,
  lineskip=2pt,
  moredelim=[is][\color{black}]{<O>}{</O>},
  moredelim=[is][\color{pragmaRed}\bfseries]{<R>}{</R>},
  moredelim=[is][\color{pragmaRed}]{<CR>}{</CR>},
  moredelim=[is][\color{pragmaGreen}\bfseries]{<G>}{</G>},
  moredelim=[is][\color{pragmaGreen}]{<CG>}{</CG>},
  moredelim=[is][\color{pragmaMagenta}\bfseries]{<M>}{</M>},
  moredelim=[is][\color{pragmaMagenta}]{<CM>}{</CM>},
  moredelim=[is][\color{pragmaBlue}\bfseries]{<B>}{</B>},
  moredelim=[is][\color{pragmaBlue}]{<CB>}{</CB>},
  moredelim=[is][\color{orange}]{<OR>}{</OR>},
  moredelim=[is][\color{orange}\bfseries]{<COR>}{</COR>},
}

\newcommand*\circled[1]{\tikz[baseline=(char.base)]{ \node[shape=circle,fill,inner sep=0.4pt] (char) {\textcolor{white}{#1}};}}

\newcommand{\framework}{MailoHLS\xspace}

\definecolor{dmblue}{HTML}{0b03fc}
\newcommand{\DM}[1]{\textcolor{dmblue}{#1}}
\definecolor{darkgreen}{RGB}{0,100,0}

\newcommand{\EV}[1]{\textcolor{orange}{#1}}

\newcommand{\addcites}[1]{\textcolor{red}{~[XXX]}}

\definecolor{lightgray}{gray}{0.95}

\lstdefinestyle{shortlststyle}{
    backgroundcolor=\color{algobg},
    basicstyle=\ttfamily\footnotesize,
    frame=tb,                
    rulecolor=\color{black},
    framerule=0.5pt,
    xleftmargin=0pt,
    xrightmargin=0pt,
    aboveskip=0.5em,
    belowskip=0.5em,
    moredelim=[is][\color{pragmaGreen}\bfseries]{<R>}{</R>},
}

\begin{document}

\title{MailoHLS: Multi-Adapter Structure-Aware Learning for Pareto-Driven HLS Pragma Optimization}


\author{Elena Vouvali}
\email{elenavouvali@gmail.com}
\affiliation{%
  \institution{National Technical University of Athens}
  \department{Department of Electrical and Computer Engineering}
  \city{Athens}
  \country{Greece}
}

\author{Dimosthenis Masouros}
\email{demo.masouros@microlab.ntua.gr}
\affiliation{%
  \institution{National Technical University of Athens}
  \department{Department of Electrical and Computer Engineering}
  \city{Athens}
  \country{Greece}
}

\author{Aggelos Ferikoglou}
\email{aferikoglou@microlab.ntua.gr}
\affiliation{%
  \institution{National Technical University of Athens}
  \department{Department of Electrical and Computer Engineering}
  \city{Athens}
  \country{Greece}
}

\author{Dimitrios Soudris}
\email{dsoudris@microlab.ntua.gr}
\affiliation{%
  \institution{National Technical University of Athens}
  \department{Department of Electrical and Computer Engineering}
  \city{Athens}
  \country{Greece}
}

\author{Sotirios Xydis}
\email{sxydis@microlab.ntua.gr}
\affiliation{%
  \institution{National Technical University of Athens}
  \department{Department of Electrical and Computer Engineering}
  \city{Athens}
  \country{Greece}
}



\begin{abstract}

High-Level Synthesis (HLS) enables rapid development of FPGA accelerators, yet achieving high-quality results (QoR) remains challenging due to the large and irregular design space induced by compiler directives (a.k.a \textit{pragmas}). Selecting effective configurations requires reasoning over complex interactions between program structure, memory behavior, and often conflicting objectives such as latency and resource utilization. 
Prior model-driven approaches exhibit limited generalization across kernels and fail to capture higher-level optimization intent. 
Recently, Large Language Models (LLMs) capture code semantics and high-level intent, but their sequential representations hinder modeling of structural dependencies and global trade-offs, leading to suboptimal HLS designs.

We present \framework, a hybrid framework that combines LLM-based semantic reasoning with GNN-based structural modeling for objective-aware directive optimization. By integrating structural embeddings via cross-attention and leveraging PEFT with objective-conditioned LoRA adapters and Pareto-driven optimization, \framework enables joint reasoning over code semantics, structure, and design trade-offs.
Across seen and unseen kernels, MailoHLS achieves up to 12.42$\times$ and 8.4$\times$ speedup (9.48$\times$ and 4.97$\times$ geometric mean) for latency optimization, consistently producing near-Pareto-optimal designs. On fully unseen applications, it reaches up to 10.2$\times$ speedup (6.58$\times$ geometric mean), outperforming high-end LLMs and prior approaches while narrowing the gap to the Pareto frontier.


\end{abstract}



\maketitle

\section{Introduction}

Hardware accelerators are increasingly deployed across cloud and edge environments, with field-programmable gate arrays (FPGAs) playing a key role due to their flexibility and energy efficiency~\cite{tomkou2026linking}. 
Cloud platforms such as Microsoft’s Project Catapult~\cite{putnam2014reconfigurable}, IBM’s CloudFPGA~\cite{bobda2022future}, and Amazon EC2 F2 instances~\cite{amazon_f1_instance} highlight their adoption in large-scale, latency-critical services. 
At the edge, FPGAs are also widely used in domains such as networking~\cite{chamola2020fpga} and robotics~\cite{wan2021survey}, where their deterministic execution, low latency, and reconfigurability suit real-time workloads.

Despite their advantages, designing efficient FPGA accelerators remains challenging.
Achieving high performance typically requires low-level hardware design in Hardware Description Languages (HDL), which is time-consuming and demands significant expertise.
High-Level Synthesis (HLS) addresses this by raising the abstraction level from HDL to C/C++-like specifications~\cite{cong2011hls}, designers guide the generated microarchitecture through compiler directives (a.k.a. \textit{pragmas}) that control loop transformations, parallelism, and memory behavior.
Consequently, achieving high-quality results (QoR), such as latency and resource utilization, critically depends on selecting effective combinations of these directives.
However, directive effects are highly interdependent: decisions on loops or memory structures can impact other parts of the program in non-obvious ways.
For example, memory partitioning directly constrains exploitable parallelism; insufficient partitioning may render aggressive loop unrolling infeasible due to bandwidth limitations.
This leads to a vast design space, where each configuration must be evaluated through costly synthesis runs (often minutes to hours). 
The problem is further compounded by the need to balance multiple, often conflicting objectives, such as latency, resource utilization, and energy efficiency.

To tackle this challenge, prior research has introduced automated design space exploration (DSE) frameworks for high-level synthesis (HLS). 
Early efforts primarily relied on \textit{synthesis-based} approaches~\cite{mahapatra2014machine, sengupta2015user, ferretti2018lattice, ferretti2018cluster}, which explore the configuration space by evaluating quality-of-result (QoR) metrics obtained from designs compiled with HLS tools. 
While these methods are generally framework-agnostic, they often suffer from slow convergence, as insights about the design space are gathered incrementally and each exploration step requires expensive synthesis runs.
More recently, \textit{model-driven}~\cite{xydis2014spirit, dai2018fast, makrani2019xppe, lin2020hl} and \textit{data-driven}~\cite{ferretti2020leveraging, ferikoglou2023collectivehls, ferikoglou2024data, ferikoglou2024collectivehls} approaches have been proposed to mitigate the cost of iterative synthesis. 
Model-driven techniques replace costly synthesis in the optimization loop with predictive models that estimate QoR metrics, whereas data-driven methods leverage knowledge from previously explored designs to guide optimization for new applications. 
However, many model-driven approaches remain application-specific, often requiring retraining for each new design.
To address this limitation, graph neural networks (GNNs) have emerged as a promising direction~\cite{ustun2020accurate, ferretti2022graph, sohrabizadeh2022automated, sohrabizadeh2023harp}. 
By representing programs as graphs, GNNs can effectively capture structural characteristics, such as control flow, data dependencies, and memory access patterns, enabling more generalizable and efficient design space exploration~\cite{murphy2024balor}.
Yet, these approaches primarily focus on structural representations and often lack the ability to reason about higher-level semantics or optimization intent. 
As a result, their effectiveness can be limited when generalizing across diverse workloads or adapting to different optimization objectives.


In parallel, recent advances in Large Language Models (LLMs) demonstrate strong capabilities in code understanding and generation, motivating their application to hardware design automation~\cite{pan2025survey}.
State-of-the-art models, such as GPT-4~\cite{openai2023gpt4} and Claude~\cite{anthropic2024claude3}, are already integrated into development environments (e.g., VS Code), enabling developers to interactively generate, refine, and optimize code within their workflows.
However, in the context of HLS optimization, LLMs often struggle to produce efficient and reliable directive configurations, particularly for complex kernels, and may even generate invalid or incorrectly placed pragmas.
The reason is that LLMs primarily rely on sequential token representation and generation without explicit modeling of program structure or non-functional objectives.
This makes it difficult to capture complex dependencies across loops, memory accesses, and hardware resources, as well as to reason about performance metrics such as latency and resource utilization.
Moreover, recent LLM-based research works typically rely on structure-agnostic adaptation techniques, such as prompt engineering~\cite{hong2024llm, fu2023gpt4aigchip}, retrieval-based augmentation~\cite{collini2025c2hlsc, xiong2024hlspilot}, or indirect incorporation of structural information~\cite{wang2025llm, prakriya2025lift}. 
These methods provide implicit guidance but do not explicitly model structural dependencies or hardware constraints. 
Hence, their ability to generate efficient and valid configurations remains limited.

In this paper, we propose \framework, a hybrid approach for HLS optimization that combines \textit{structural} and \textit{semantic} reasoning.
To capture structural information, \framework constructs a directive-aware graph representation encoding control flow, data dependencies, and memory interactions.
In parallel, it leverages an LLM backbone to model source-level semantics and optimization intent.
These representations are fused through \textit{cross-attention}, enabling joint reasoning over structural constraints and semantic context.
Rather than relying on code generation, \framework reformulates HLS optimization as a directive-level prediction problem, where the model assigns values to predefined directive placeholders.
This design (i) constrains the search space to valid configurations, (ii) improves robustness by focusing on optimization decisions rather than syntax generation, and (iii) aligns with modern HLS workflows.
Finally, \framework incorporates objective-aware adaptation through lightweight LoRA adapters, enabling specialization across different design goals.
More specifically, MailoHLS introduces:
\begin{itemize}[leftmargin=*, topsep=0pt]
    \item \textbf{Hybrid structural–semantic modeling for HLS optimization:}
    We propose \framework, which combines directive-aware graph representations with LLM-based semantic reasoning via cross-attention, enabling joint reasoning over program structure and optimization decisions.

    \item \textbf{Directive-level formulation of HLS optimization:}
    Instead of relying on LLM-based code generation or modification, we reformulate HLS optimization as a structured prediction problem over directive placeholders, decoupling optimization decisions from code synthesis and constraining the search space to valid configurations.

    \item \textbf{Pareto-driven, objective-aware optimization:}
    We introduce a training framework based on Pareto-ranked design data and lightweight LoRA adapters, leveraging supervised fine-tuning and preference-based optimization to specialize the model across multiple objectives.
\end{itemize}

We evaluate MailoHLS on seen kernels, leave-one-family-out transfer, and fully unseen applications. On seen kernels, it achieves 9.48×, 7.95×, and 2.10$\times$ geometric-mean speedup for latency-, balanced-, and resource-oriented objectives, closely tracking—and often surpassing—Pareto reference points. Under the MachSuite holdout setting, it retains strong performance (4.97$\times$, 5.21$\times$, 2.45$\times$), demonstrating robust generalization across unseen kernel families. 
On fully unseen applications, it achieves up to 6.58$\times$ speedup, outperforming high-end LLMs, which frequently produce invalid or suboptimal designs, while consistently generating Pareto-efficient and synthesizable configurations.

\section{Background \& Related Work}

\subsection{HLS Directive Optimization}

\begin{lstlisting}[style=hlsalgo, float=t,
caption={GEMV HLS kernel with different pragma configurations.},
label={alg:gemv-hls}]
<O>const float</O> <CR>A[N][M]</CR>;
<R>#pragma HLS ARRAY_PARTITION variable=A cyclic factor=8 dim=2</R>
<O>const float</O> <CB>x[M]</CB>;
<B>#pragma HLS ARRAY_PARTITION variable=x complete</B>
<O>float</O> <OR>y[N]</OR>;
<COR>#pragma HLS ARRAY_PARTITION variable=y complete</COR>
<CG>for</CG> <O>(int i = 0; i < N; i++) {</O>
<G>#pragma HLS PIPELINE II=1</G>
  <O>float acc = 0;</O>
  <CM>for</CM> <O>(int j = 0; j < M; j++) {</O>
    <M>#pragma HLS UNROLL factor=8</M>
    acc += A[i][j] * x[j];
  }
  y[i] = acc;
}
\end{lstlisting}

HLS is an FPGA design methodology that translates high-level programming descriptions into hardware accelerators, reducing the need for manual HDL design~\cite{rupnow2011study}. 
By abstracting low-level implementation details, HLS allows designers to focus on functional behavior, while the compiler performs scheduling, resource allocation, and Register Transfer Level (RTL) generation~\cite{cong2011high}. 
To optimize performance and resource efficiency, developers use directives or \textit{pragmas} to guide the compiler, enabling microarchitectural customization and exploration of design trade-offs. 
These directives shape key aspects of the design, such as parallelism, pipelining, and memory behavior, that directly impact QoR metrics, including latency and resource utilization~\cite{papakonstantinou2011multilevel}.

Listing~\ref{alg:gemv-hls} shows the HLS code for a General Matrix–Vector Multiplication (GEMV) kernel. 
The first directive applies cyclic partitioning to array \textit{A} across $8$ memory banks along the second dimension, enabling concurrent column accesses and increasing memory bandwidth. 
The unrolling directive replicates the inner loop by a factor of $8$, allowing multiple multiply–accumulate operations to execute in parallel and improving throughput. 
Each directive is associated with a specific source-code location (Action Point -- AP), which affects the generated hardware. 
For example, loop-level directives are applied immediately after the corresponding loop declaration.

The effectiveness of HLS highly depends on selecting appropriate pragmas for each application. 
However, identifying optimal configurations for a given objective remains challenging~\cite{schafer2019high}, due to the large design space and diverse directive options~\cite{numan2020towards}. 
For example, as shown in Listing~\ref{alg:gemv-hls}, array partitioning supports multiple strategies (e.g., \textit{block}, \textit{cyclic}, \textit{complete}) with tunable factors across dimensions. 
Achieving high performance requires coordinated use of multiple directives~\cite{ferikoglou2024data, ferikoglou2024collectivehls}, as poor choices can lead to inefficient resource utilization and underutilized parallelism, sometimes resulting in performance below that of CPUs~\cite{chi2022democratizing}. 


\noindent\textbf{Automating Directive Optimization:} To alleviate manual tuning, prior work has explored automated design space exploration (DSE) for HLS directive optimization. 
However, this problem remains challenging due to the exponential growth of the design space with the number of action points and directive options, the high cost of evaluating each configuration via full HLS synthesis, and the need to balance multiple, often conflicting objectives, such as latency or allocated resources on the device.
Early \textit{synthesis-based} methods~\cite{mahapatra2014machine, 10.1145/2390191.2390202, ferretti2018lattice, ferretti2018cluster} treat the HLS tool as a black box, iteratively querying it to obtain QoR metrics. While general, they suffer from slow convergence due to expensive evaluations. To mitigate this, \textit{data-driven} approaches~\cite{ferretti2020leveraging, ferretti2021db4hls, bai2023towards, ferikoglou2023collectivehls, ferikoglou2024data, ferikoglou2024collectivehls, ferikoglou2026gnomegasis} reuse prior design knowledge to guide exploration, reducing the number of required synthesis runs. 
In parallel, early \textit{model-driven} methods~\cite{xydis2014spirit, dai2018fast, makrani2019xppe, lin2020hl} employed ANN and CNN models to predict QoR metrics from pre-synthesis features, avoiding full compilation.
Despite these advances, existing approaches either rely on costly evaluations or depend on handcrafted features and limited generalization.

\subsection{Graph Neural Networks}
\label{subsec:gnn-background}


GNNs have been widely adopted as an alternative for HLS optimization to model and explore the large design space induced by pragma configurations~\cite{gao2024hierarchical, bai2024learning, li2025hierarchical, murphy2024balor, wu2022hls, sohrabizadeh2023harp, ustun2020accurate, sohrabizadeh2022automated, bai2022improving}. 
As shown in Figure~\ref{fig:background:gnn}, they learn structure-aware representations through iterative message passing, where each node aggregates information from its neighbors and applies a learnable transformation, capturing progressively broader dependencies.
Thus, \textit{GNNs learn structure-aware representations that capture graph topology and dependencies}.

GNNs align naturally with HLS, where kernels are represented as graphs (e.g., Control Data Flow Graphs -- CDFGs) encoding control and data dependencies. 
However, QoR often depends on long-range interactions (e.g., between nested loop transformations and memory hierarchy), which are difficult to capture with standard graph representations and localized message passing. Modeling such dependencies requires deep GNNs, which can suffer from over-smoothing and training instability~\cite{alon2021bottleneck,oono2020graph,xu2019powerful}. 
To address this, approaches such as HARP~\cite{sohrabizadeh2023harp} augment graphs with \textit{auxiliary nodes} that encode higher-level structure, shortening distances between dependent nodes and improving information propagation.

\subsection{Transformer, Cross-Attention and LoRA}
\label{subsec:llm-background}
While GNNs effectively capture structural dependencies, they struggle to model code semantics and global optimization intent. Recent advances in LLMs have significantly improved their ability to reason about code and perform complex program transformations, driving growing interest in hardware design automation~\cite{masouros2024lbr,prakriya2025lift,qin2024crossmodality,hong2024llm,swaroopa2025evaluating,collini2025c2hlsc}. 
At the same time, LLMs such as GPT-4~\cite{openai2023gpt4} and Claude~\cite{anthropic2024claude3} are already integrated into development environments (e.g., VSCode), enabling interactive code generation and refinement.

\begin{figure}[t]
    \centering

    \begin{subfigure}[t]{0.29\linewidth}
        \centering
        \includegraphics[width=\linewidth]{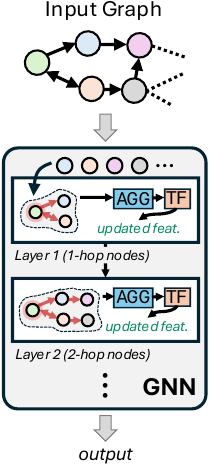}
        \caption{GNN}
        \label{fig:background:gnn}
    \end{subfigure}
    \hfill
    \begin{subfigure}[t]{0.67\linewidth}
        \centering
        \includegraphics[width=\linewidth]{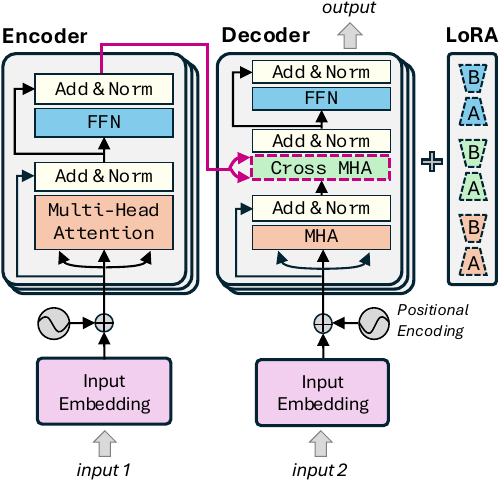}
        \caption{Transformer and Low-Rank Adaptation}
        \label{fig:background:llm}
    \end{subfigure}
    
    \caption{Overview of GNN and Transformer architectures.}
    \label{fig:llm-lora-gnn-background}
\end{figure}

Transformers~\cite{vaswani2017attention}, the backbone of LLMs, learn context-aware semantic representations by modeling relationships across tokens through attention mechanisms. 
As shown in Figure~\ref{fig:background:llm}, they can be structured as encoder–decoder architectures, where attention (MHA) enables global information exchange, while feed-forward layers (FFN) apply learned token-level transformations. 
Cross-attention (cross MHA) further enables the integration of external representations, making it well-suited for fusing complementary modalities such as structural features.
Thus, \textit{Transformers learn context-aware semantic representations that capture relationships across tokens and long-range dependencies in the input.}

Adapting large Transformer models to domain-specific tasks is computationally expensive due to their scale. Parameter-Efficient Fine-Tuning (PEFT) addresses this by updating only a small subset of parameters. Low-Rank Adaptation (LoRA)~\cite{hu2022lora} injects lightweight trainable updates into selected layers, enabling efficient specialization. In HLS, such adapters can be conditioned on different optimization objectives, allowing the model to balance competing design goals while preserving general knowledge.

\section{Motivation}
\label{sec:motivation}

\subsection{Structural nature of HLS optimization}
\label{subsec:hls-structural}

HLS optimization is inherently structural, as performance depends not only on selecting and placing directives, but also on their interactions, which collectively shape the generated hardware. Directives influence key aspects such as parallelism and memory access, and are tightly interdependent. For example, array partitioning determines available memory bandwidth, which constrains achievable parallelism through loop unrolling. In Listing~\ref{alg:gemv-hls}, partitioning arrays \textit{A} and \textit{x} with a factor of $2$ makes an inner loop unroll factor of $8$ infeasible, as the memory system cannot supply operands fast enough to sustain parallel execution. Such imbalances lead to resource underutilization and degraded performance. Consequently, HLS optimization requires reasoning over a tightly coupled design space, where QoR is determined by the global hardware organization rather than isolated directive choices, motivating models that capture structural dependencies and their interactions.


\subsection{Limitations of Decoder-Only LLMs for HLS}
\label{subsec:llm-limitations}

\begin{figure}[t]
    \centering
    \includegraphics[width=\linewidth]{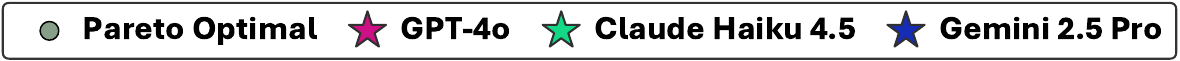}
    \\
    \vspace{2pt}
    \begin{subfigure}[b]{0.48\linewidth}  
        \centering
        \includegraphics[width=\linewidth]{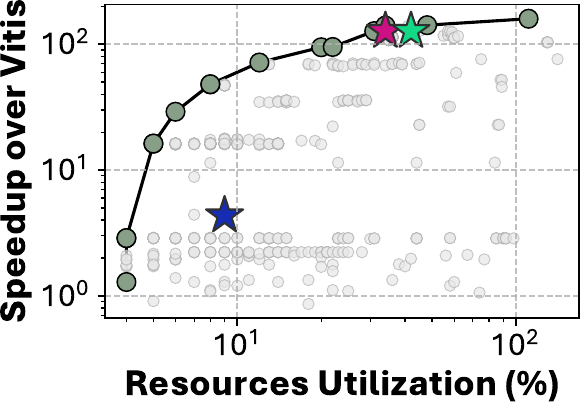}
        \caption{GEMV}
        \label{fig:gemv-pf}
    \end{subfigure}
    \hfill
    \begin{subfigure}[b]{0.48\linewidth}  
        \centering
        \includegraphics[width=\linewidth]{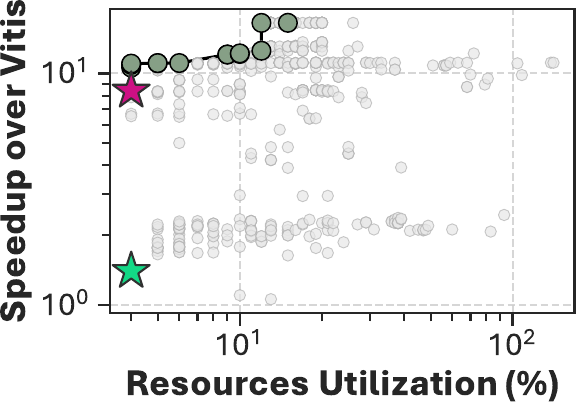}
        \caption{KNN}
        \label{fig:knn-pf}
    \end{subfigure}
    \vspace{-5pt}  
    \caption{Positioning of decoder-only LLM generated designs compared to Pareto-optimal configurations from ~\cite{ferikoglou2026gnomegasis}.}
    \label{fig:llm-decoder-limitation}
\end{figure}

Modern large-scale LLMs, such as GPT-4o~\cite{openai2023gpt4} and Claude Haiku 4.5~\cite{anthropic2024claude3}, follow a decoder-only architecture, modeling code as a sequence of tokens and capturing semantic relationships through autoregressive attention. 
While this enables strong capabilities in code understanding and generation~\cite{chang2024survey}, it inherently limits their ability to explicitly model structural dependencies in programs.
However, as discussed in \S\ref{subsec:hls-structural}, in HLS performance optimization is fundamentally driven by structural transformations and their interactions. 
Although LLMs benefit from strong generalization and extensive training on large code corpora, their token-centric reasoning often fails to capture hardware-level constraints and dependencies required for effective directive optimization.


To systematically investigate these limitations, we evaluate the capability of decoder-only LLMs to optimize HLS designs through prompt-based interaction. 
We consider three applications of increasing complexity: (i) the GEMV kernel of Listing~\ref{alg:gemv-hls}, (ii) a K-Nearest Neighbors (KNN) implementation from the RodiniaHLS benchmark suite~\cite{cong2018understanding}, and (iii) a Kalman filtering application sourced from GitHub.
The LLMs are instructed to minimize latency under resource constraints on an AMD UltraScale+ ZCU104 platform (100 MHz), restricted to pragma insertion without modifying the source code.
We evaluate GPT-4o, Claude Haiku 2.5, and Gemini 2.5 Pro. Generated designs are synthesized with Vitis HLS 2021.1 and assessed in terms of latency speedup and resource utilization against a baseline without directives (Vitis). To contextualize performance, we derive Pareto-optimal configurations using an NSGA-II-based design space exploration framework~\cite{ferikoglou2026gnomegasis}.

Figure~\ref{fig:llm-decoder-limitation} shows the Pareto frontiers obtained via NSGA-II and the LLM-generated designs for GEMV and KNN.
For GEMV, the latency-efficient Pareto design achieves a $159.5\times$ speedup over baseline Vitis, using 1\%, 11\%, 34\%, and 65\% of BRAM, DSP, FF, and LUT, respectively. 
In contrast, Gemini 2.5 Pro attains only $4.4\times$, failing to identify the \textit{y} array as critical for throughput. 
As a result, no partitioning is applied, creating a memory bottleneck that prevents achieving an initiation interval (II) of 1 and limits the benefits of loop unrolling.
In contrast, GPT-4o and Claude Haiku 2.5 correctly identify \textit{y} as performance-critical and apply appropriate partitioning, enabling efficient memory access and sustained pipeline throughput, achieving up to $127.2\times$ speedup within resource constraints.

Regarding KNN, the latency-efficient Pareto design achieves a $17\times$ speedup over baseline Vitis, while utilizing only 1\%, 1\%, 2\%, and 11\% of BRAM, DSP, FF, and LUT resources, respectively. In contrast, Gemini 2.5 Pro generates directives that result in a compilation failure. 
Specifically, it introduces dataflow without properly restructuring memory accesses, causing multiple processes to read concurrently from the same global memory interface and violating the single-port constraints of bundled AXI interfaces in Vitis HLS.
This behavior highlights a key limitation of general-purpose LLMs, which often fail to account for hardware-level constraints.
A similar issue is observed with Claude Haiku 2.5: although loop unrolling is applied, the corresponding arrays are not partitioned, preventing true parallel memory access and limiting performance to $1.4\times$ speedup. In contrast, GPT-4o adopts a more balanced strategy by combining pipelining (II=1), dataflow, and proper memory management, achieving a higher speedup of $8.4\times$, which nevertheless remains well below the optimal achievable performance.

Finally, for the Kalman filter, which is the most complex application considered, the latency-efficient Pareto design achieves a $35.3\times$ speedup over baseline Vitis, with resource utilization of 92\%, 7\%, 7\%, and 19\% for BRAM, DSP, FF, and LUT, respectively.
In this case, GPT-4o generates directives that lead to compilation failure due to misplaced array pragmas, applied before their corresponding declarations, whichi result in undefined variable errors.
For Claude Haiku 2.5 and Gemini 2.5 Pro, both models apply overly aggressive directives, leading to BRAM over-utilization of up to 113\%. \textit{This behavior highlights the difficulty for general-purpose LLMs to consistently satisfy hardware resource constraints, often leading to impractical designs.}

\subsection{Pitfalls of Structure-Agnostic Adaptation}

High-level adaptation techniques, such as prompt engineering~\cite{pan2025survey} and retrieval-augmented generation (RAG)~\cite{lewis2020retrieval}, aim to guide LLMs through additional context and structured instructions. 
This concept has also been investigated in the HLS domain~\cite{xiong2024hlspilot}, where enriched prompts explicitly specify which directives should be applied and where they should be placed within the code.
While these approaches improve guidance, they remain fundamentally structure-agnostic, relying on token-level context rather than explicit representations of program structure.

To assess their effectiveness, we design prompts that encode key HLS optimization strategies, including pipelining, loop unrolling, and array partitioning~\cite{ferretti2020leveraging}. Beyond directive placement, the prompts also include tuning guidance (e.g., enforcing II=1 or aligning partitioning factors with loop unrolling). We evaluate this approach on KNN and Kalman filtering, representing the more complex applications.
For KNN, enriched prompts improve performance across models. In particular, Gemini 2.5 Pro generates a valid design by correctly applying dataflow under AXI constraints, achieving an $8.4\times$ speedup over baseline. 
GPT-4o and Claude Haiku 4.5 also benefit, reaching $6\times$ and $15.6\times$ speedups, respectively.
However, this improvement does not generalize to more complex cases. In the Kalman filter, GPT-4o still produces invalid code due to incorrect pragma placement, while Claude Haiku 2.5 and Gemini 2.5 Pro exceed BRAM capacity (up to 113\%), despite explicit guidance.

These results show that, \textit{even with detailed and structured prompts, general-purpose LLMs struggle to consistently generate valid and efficient HLS designs}. While prompt engineering can guide the model toward better local decisions, it does not provide an explicit representation of program structure or hardware constraints. As a result, the model must infer complex structural relationships implicitly from token sequences, which is often unreliable, especially for long-range dependencies and tightly coupled optimizations. This limitation becomes more pronounced as application complexity increases, where multiple interacting transformations must be coordinated. Consequently, structure-agnostic adaptation remains insufficient for reliably navigating the HLS design space.
\section{\framework: Design Overview}

\begin{figure*}[t]
    \centering
    \includegraphics[width=\textwidth]{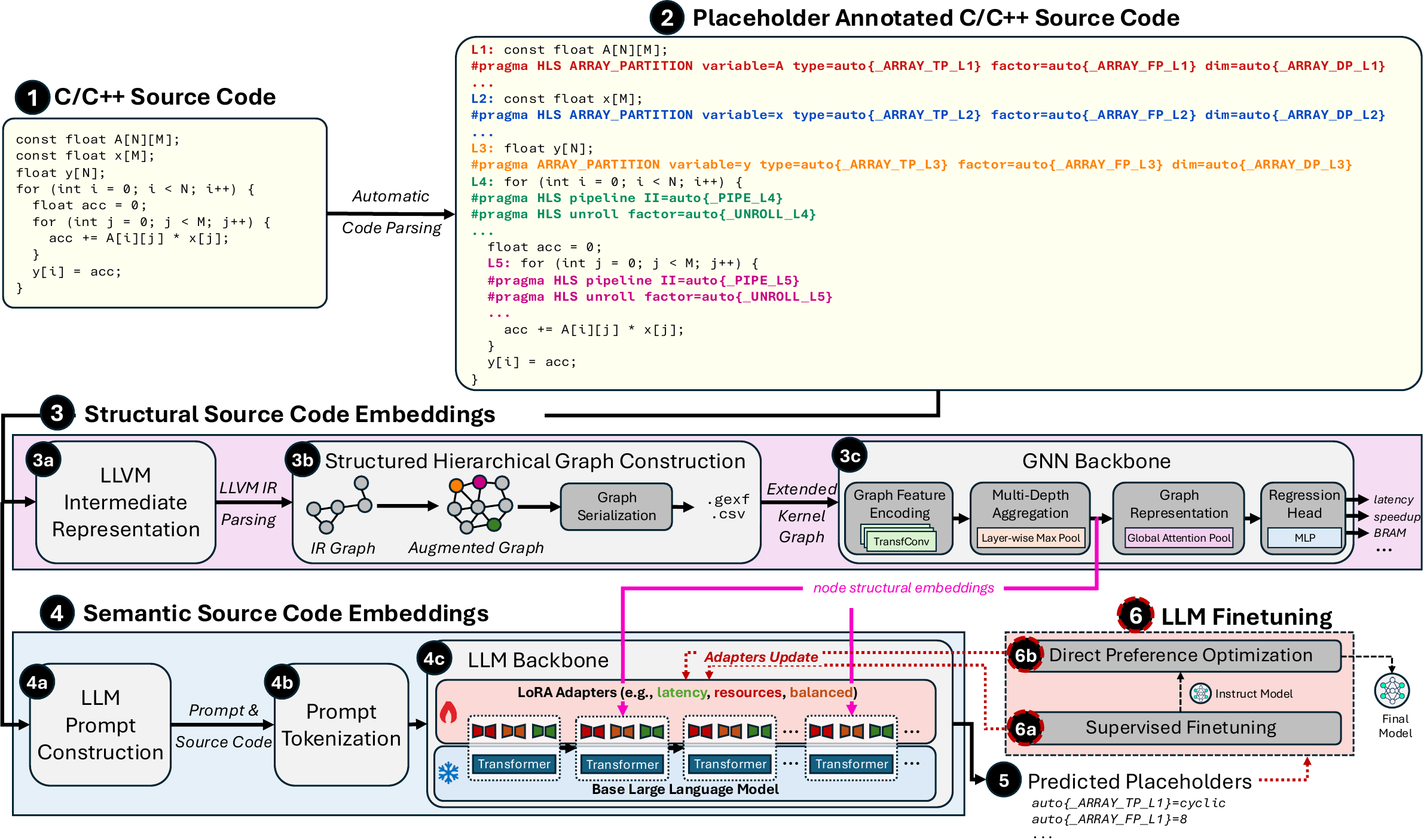}
    \caption{\framework Design Overview. Given a C/C++ kernel, \framework inserts directive placeholders and encodes structural and semantic embeddings via a GNN and an LLM, fused through cross-attention, with objective-specific LoRA adapters. Cold (\includegraphics[height=.8em]{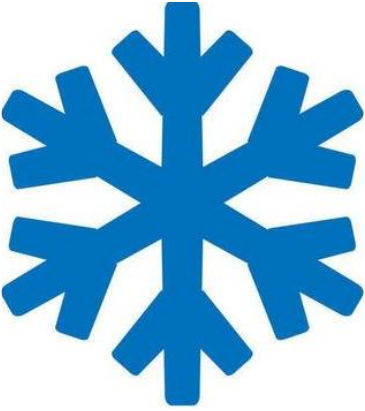}) components denote frozen modules, while hot (\includegraphics[height=.8em]{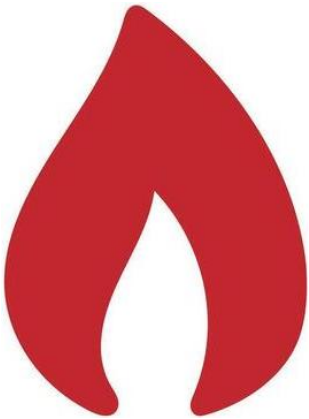}) components denote trainable parts.}
    \label{fig:design-overview}
\end{figure*}

To address these limitations, we propose \framework, a Pareto-driven hybrid framework for HLS pragma optimization that explicitly integrates structural reasoning, semantic understanding and objective-aware adaptation.
\framework is designed around three principles:

\begin{enumerate}[leftmargin=*, label=\textbf{(\arabic*)}, labelsep=0.3em,topsep=0pt]
    \item \textbf{Joint structural \& semantic modeling.} \framework combines graph-based structural representations with language-model reasoning to capture both the program organization and its high-level semantics. Unlike text-only approaches, it explicitly models interactions between loop transformations, memory accesses, and directive placement, enabling the model to reason about hardware-relevant constraints and their impact on QoR.
    \item \textbf{Directive-Level Decision Abstraction.} Instead of training the LLM to generate or modify code, \framework formulates HLS optimization as a structured prediction problem over directive placeholders inserted at valid action points (e.g., loops and arrays). This abstraction decouples where directives apply from what values to assign, constraining the search space to valid configurations and improving robustness.
    \item \textbf{Objective-Aware Optimization.} \framework supports multiple optimization goals (e.g., latency, area, or balanced trade-offs) through objective-conditioned specialization. By leveraging Pareto-ranked design data and preference-based alignment, the model learns to navigate trade-offs between competing objectives, enabling it to generate objective-aware configurations.
\end{enumerate}
    
Figure~\ref{fig:design-overview} presents the end-to-end workflow of \framework. Given an input C/C++ kernel (\circled{1}), the framework first identifies candidate optimization action points (e.g., loops and arrays) and augments them with parameterized directive placeholders (\circled{2}), which define the decision space for pragma selection.
\framework then constructs two complementary representations of the annotated program. 

For the \textit{structural representation} (\circled{3}), the code is translated into an extended hierarchical graph derived from LLVM IR, enriched with auxiliary nodes and directive-aware connections that capture loop hierarchy, memory behavior, and long-range dependencies. A GNN encoder processes this graph to produce node embeddings trained to reflect QoR-relevant properties (e.g., parallelism, data reuse, and resource pressure). 
These embeddings are aligned with directive placeholders and injected into the LLM through \textit{cross-attention}.

For the \textit{semantic representation} (\circled{4}), \framework employs a decoder-only LLM to model code semantics and directive intent. To integrate structural context, the placeholder-aligned graph embeddings are injected into the LLM through cross-attention layers, applied selectively at directive prediction tokens. A gating mechanism regulates the contribution of structural and semantic signals, allowing the model to jointly reason about program logic and hardware constraints when predicting pragma values.

Finally, \framework predicts the values of the directive placeholders (\circled{5}) by jointly leveraging semantic and structural information. To support different optimization goals, it enables objective-aware specialization through lightweight adapter modules (\circled{6}). Each adapter is trained for a specific objective using supervised fine-tuning followed by preference-based alignment on Pareto-ranked design pairs. At inference time, the appropriate adapter is activated based on the target objective, guiding the model toward configurations that reflect the desired trade-off.
\section{\framework: Detailed Design}


\subsection{Code Parsing \& Placeholder Placement}
To avoid syntactic inconsistencies (\S\ref{subsec:llm-limitations}), \framework operates over a structured formulation of the optimization problem. 
Given an input kernel \circled{1}, \framework first automatically parses the code to identify candidate action points where optimization pragmas can be applied, such as loop constructs and array definitions. 
These locations define the target design space and directly correspond to performance–resource trade-offs in HLS.
Each action point is annotated with a directive placeholder token (e.g., \texttt{<L1>, <L2>, <L3>)} and every placeholder is associated with a predefined set of valid, yet unassigned, pragmas (\circled{2}). 
For example, for a loop construct, this set may include pipelining, loop unrolling, tiling and other transformations, as follows:
\begin{lstlisting}[style=shortlststyle]
<R>L4:</R> for (int i = 0; i < N; i++) {
#pragma HLS pipeline II=<R>auto{_PIPE_L4}</R>
#pragma HLS unroll factor=<R>auto{_UNROLL_L4}</R>
...
\end{lstlisting}

\noindent This representation converts the kernel into a templated program where directive placement is fixed and only their values remain to be determined by the LLM backbone (\S\ref{subsec:llm-backbone}). 
As a result, the model is restricted to operate over valid transformation points and predict directive values for each placeholder.

\subsection{Structural Source Code Embeddings}
After annotating the code, \framework extracts \textit{structural embeddings} by lowering the input kernel to an intermediate representation~\cite{lattner2004llvm}, building a hierarchical graph (\circled{\footnotesize3b}) and encoding it with a GNN (\circled{\footnotesize3c}).

\subsubsection{LLVM Intermediate Representation}
To enable consistent structural analysis across kernels, \framework leverages an intermediate representation (IR) of the input code.
IRs provide a normalized view of control flow, data dependencies, and memory operations, abstracting away syntactic variability in the code, such as variable naming and equivalent loop formulations~\cite{sohrabizadeh2023harp, ye2022scalehls, basalama2025streamhls}.
Specifically, \framework lowers the input kernel to LLVM IR~\cite{lattner2004llvm}, a low-level, Static Single Assignment (SSA)-based compiler representation.
LLVM IR exposes constructs that directly relate to hardware behavior.
For example, loop guards are represented through compare instructions (e.g., \texttt{icmp}), control transitions through branch instructions (e.g., \texttt{br}), and value and memory dependencies through explicit dataflow.
A simplified example is shown below:

\begin{lstlisting}[style=shortlststyle]
...
for.cond:
  %0 = load i32, i32* %i, align 4
  %cmp = icmp slt i32 %0, 32
  br i1 %cmp, label %for.body, label %for.end12
...
\end{lstlisting}

From this representation, \framework identifies loop constructs by tracing compare instructions (e.g., \texttt{icmp}) and their associated branch targets (e.g., \texttt{br}), grouping them into loop-level regions that capture both control structure and associated computation.
These structures expose hardware-relevant constraints such as pipelining feasibility, scheduling dependencies, and available parallelism.
The resulting IR is then used as the basis for structured graph construction (\S\ref{subsubsec:graph-construction}) and subsequent GNN-based encoding (\S\ref{subsubsec:gnn-backbone}) to produce structural embeddings.

\subsubsection{Structured Hierarchical Graph Construction}
\label{subsubsec:graph-construction}
Given the LLVM IR, \framework constructs a ProGraML~\cite{cummins2021programl}-style graph representation, which encodes instructions, values, and control/data dependencies as a unified, SSA-based graph, that can be directly consumed by graph neural networks.
However, directly applying such a representation to HLS optimization presents two key challenges.
First, directive positions are not encoded in the IR, and therefore optimization pragmas that control parallelism and resource utilization are not explicitly represented.
Second, the graph may exhibit long-range dependencies, where decisions at one loop level affect scheduling, pipelining, and resource sharing across distant regions, limiting the effectiveness of standard GNN models (\S\ref{subsec:gnn-background}).

To address these challenges, \framework adopts a hierarchical graph construction similar to HARP~\cite{sohrabizadeh2023harp}, extending the base ProGraML graph with directive-aware and hierarchical augmentations tailored for HLS.
The construction proceeds in multiple stages. 
Starting from the ProGraML graph, \framework identifies directive placeholder locations and augments the graph with additional nodes and edges to encode both optimization decisions and structural hierarchy.
Similar to HARP, \framework introduces two types of nodes: \textit{pragma} nodes and \textit{auxiliary} nodes. 
Pragma nodes encode optimization directives, while auxiliary nodes capture hierarchical structure and connect distant regions of the graph. 
\textit{Loop-related pragma nodes} (e.g., pipelining or unrolling) are attached to loop anchors in the IR, identified through compare instructions (e.g., icmp) that define loop boundaries.
Beyond loop-centric modeling, \framework extends HARP's representation with explicit support for shared data structures.
\textit{Array-related pragma nodes} (e.g., partitioning) are identified based on variable names and connected to IR nodes corresponding to array declarations and uses, with preference for declaration sites. This enables the graph to explicitly capture which compute regions interact through the same data structure.

Auxiliary nodes are then introduced to capture hierarchical scope. \textit{Loop-level auxiliary nodes} group operations within the same loop region, while \textit{data-centric auxiliary nodes} group accesses to the same array. 
These nodes are connected to their associated IR nodes and pragma nodes, and are further interconnected to form a hierarchical skeleton that reduces long-range dependencies.
This construction exposes both control-flow and data-sharing relationships to the GNN. 
In particular, shared data structures introduce implicit coupling between distant compute regions, enabling the model to capture memory-related effects such as bandwidth constraints, access parallelism, and contention.

\begin{figure}[t]
    \centering
    \includegraphics[width=1\columnwidth]{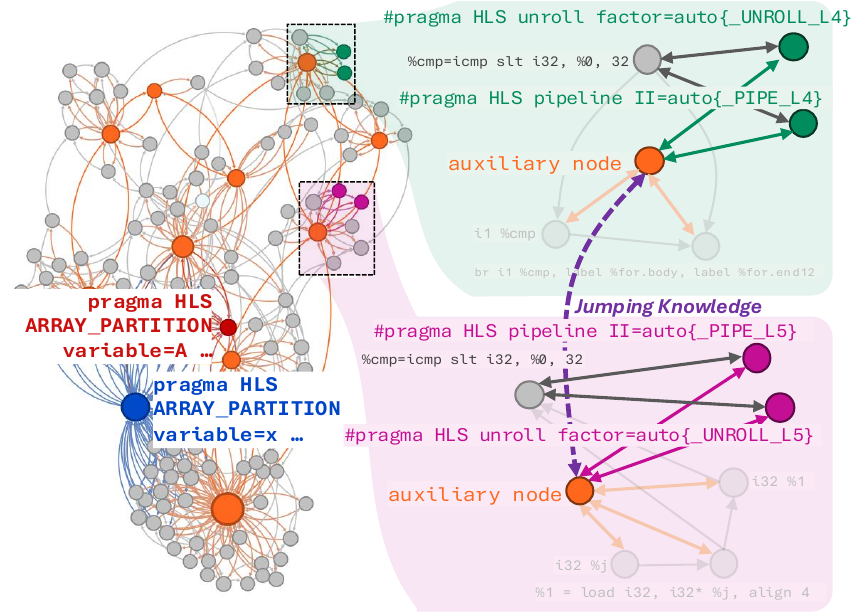}
    \caption{Graph representation of the GEMV kernel. Gray nodes represent LLVM IR operations, while colored ones show auxiliary and pragma nodes.}

    \label{fig:gemv-graph}
\end{figure}

Figure~\ref{fig:gemv-graph} shows the constructed graph for the GEMV kernel in Listing~\ref{alg:gemv-hls}.
Gray nodes represent the graph obtained directly from LLVM IR, while colored nodes represent additional auxiliary and pragma nodes.
As shown, \textit{pragma nodes} encode optimization directives and are grouped into data-centric nodes (blue/red), associated with array transformations, and control-centric nodes (green/magenta), associated with loop transformations.
Each pragma node is connected both to the corresponding LLVM IR instruction (e.g., \texttt{\%cmp} defining loop boundaries) and to its associated auxiliary node, capturing both control-flow context and data dependencies between computation and memory accesses, while enabling a direct alignment with directive placeholder prediction.
This maps directive choices in the local computation while propagating their impact across higher-level program structure.
\textit{Auxiliary nodes} (orange) capture higher-level structure and connect distant regions of the graph. Loop-level auxiliary nodes link control-flow constructs across nested loops, while data-centric auxiliary nodes aggregate accesses to shared arrays, strengthening dependencies between computation and memory behavior.

\subsubsection{GNN Backbone}
\label{subsubsec:gnn-backbone}

\framework uses a separately trained GNN to generate structural embeddings of the input kernel. 
Starting from the extended hierarchical graph (\S\ref{subsubsec:graph-construction}), the GNN applies edge-aware TransformerConv message passing to encode control-flow, data dependencies, loop structure, and memory interactions into node-level representations. 
To capture both local and long-range dependencies, we employ multi-depth aggregation via Jumping Knowledge~\cite{xu2018jumping}, which combines node embeddings from multiple GNN layers, effectively aggregating information across different hop distances.
The resulting node embeddings are further pooled using global attention to obtain a graph-level representation. 
A regression head is trained on top of these embeddings to predict HLS QoR targets (e.g., latency and resource utilization), making the GNN an independent structural surrogate prior to any LLM-based fine-tuning.
Unlike prior approaches such as HARP, \framework does not use the GNN solely as a predictor. 
Instead, it reuses the learned embeddings before the regression head as structural inputs for the LLM (\S\ref{subsubsec:graph-construction}). 
n particular, the final node embeddings are extracted and converted into placeholder-aligned memory slots, which are later consumed by the LLM cross-attention layers. 


\subsection{Semantic Embeddings \& Structural Fusion}
\label{subsec:llm-backbone}
\begin{figure}[t]
    \centering
    \includegraphics[width=1\columnwidth]{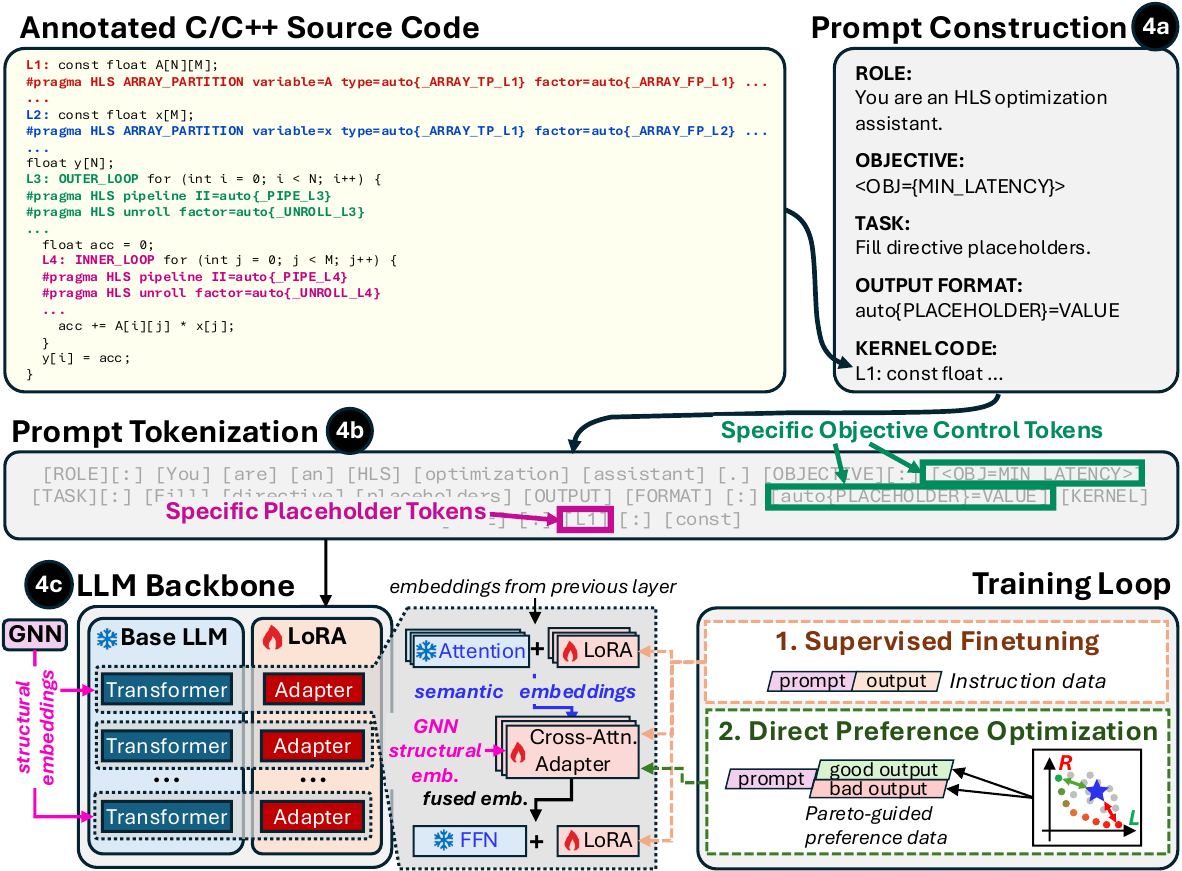}
    \caption{LLM backbone training pipeline with SFT, Pareto-guided preference optimization, and LoRA adapters.}

    \label{fig:llm-path}
\end{figure}

The core of \framework is its LLM backbone, which extracts \textit{semantic embeddings} from the source code and integrates structural embeddings from the GNN to guide directive prediction. Semantic embeddings capture higher-level program properties (e.g., access patterns and loop roles) that are not explicitly represented in the structural graph but are critical for optimization decisions.
As shown in Figure~\ref{fig:llm-path}, the annotated kernel is transformed into a structured prompt (\circled{\footnotesize4a}), tokenized (\circled{\footnotesize4b}) and processed by the LLM (\circled{\footnotesize4c}). Structural embeddings are incorporated through \textit{cross-attention layers} injected at specific blocks in the LLM.
To support multiple optimization goals, \framework adopts a multi-adapter design, where each objective is handled by a dedicated LoRA adapter.



\subsubsection{LLM Prompt Construction \& Tokenization}
\label{subsubsec:prompt}
The input source code is transformed into a structured prompt that constrains the task to directive prediction~\cite{raffel2019t5}. 
As shown at the top of Figure~\ref{fig:llm-path}, the prompt includes a role description, a target objective, a task specification, and the expected output format.
To explicitly encode this information, \framework augments the tokenizer with \textit{objective control} tokens (e.g., \texttt{<OBJ=MIN\_LATENCY>}) and \textit{placeholder} tokens.
Directive sites are represented using placeholders (e.g., \texttt{<Lk>}), while directive assignments follow a fixed schema (e.g., \texttt{auto\{PLACEHOLDER\}=VALUE}), ensuring directive locations and objectives remain atomic during tokenization.
This design enables the LLM to associate each directive decision with both its program context and optimization objective. 
In particular, placeholder tokens act as anchors that attend to (i) surrounding code regions (e.g., loop structure, memory accesses) and (ii) the objective control token, allowing each prediction to be conditioned on both structural context and the desired goal. 
As a result, the model learns a mapping from \{\textit{program context}, \textit{objective}\} to directive values, improving conditioning and task alignment~\cite{keshar2019ctrl,wang2021codet5}.
Since the directive schema is fixed, the model predicts only directive values rather than their placement, reducing the output space and improving reliability.

\subsubsection{LLM Backbone \& Structural Embeddings Fusion}
\label{subsubsec:llm-backbone}
\framework' LLM backbone is based on a decoder-only model (e.g., DeepSeek-Coder~\cite{guo2024deepseekcoder}), which encodes the semantic context of the program and predicts directive values based on the input code and the target objective. 
However, as discussed in \S\ref{subsec:llm-limitations}, semantic information alone is insufficient for HLS, as directive decisions depend on structural properties such as loop-carried dependencies, available parallelism, and memory access patterns.
Still, standard decoder-only models do not natively support attending to external representations. 

To incorporate structural information, \framework augments the backbone with \textit{cross-attention layers} that fuse GNN-derived embeddings into the token stream.
As shown in Fig.~\ref{fig:llm-path}, these layers are inserted between the self-attention module and feed-forward sublayers of selected transformer blocks, enabling structural information to be injected after semantic context aggregation and before value transformation (c.f. \S\ref{subsec:llm-background}).
To balance expressiveness and efficiency, cross-attention is applied periodically every $n$ layers (e.g., every 2--4 layers), rather than in every Tranformer block.

The structural context is derived from the GNN as a set of node embeddings, which are organized into a fixed-size structural memory.
During fusion, token embeddings act as queries ($Q$), while structural embeddings serve as keys ($K$) and values ($V$).
Crucially, alignment between tokens and structure is achieved through the directive-level abstraction: target-anchor tokens (e.g., \texttt{<Lk>}) act as routing points that attend to corresponding structural elements associated with each directive site (e.g., loop nests, memory objects).
Moreover, cross-attention is selectively applied only to target tokens corresponding to directive predictions, rather than uniformly across all tokens.
This ensures that structural information directly influences optimization decisions, while preserving the stability of the fixed output schema.
Finally, the cross-attention outputs are integrated through a gated mechanism, allowing the model to modulate the contribution of structural versus semantic signals.
This fusion enables pragma predictions to jointly account for program semantics and structural dependencies, capturing constraints that affect pipelining, scheduling, and memory behavior.

\noindent\textbf{LLM backbone sensitivity:} To assess the impact of backbone capacity on final HLS quality, we compare MailoHLS with DeepSeek-Coder-1.3B and DeepSeek-Coder-7B as the semantic backbone. As shown in Figure 6, moving from the 1.3B to the 7B backbone increases the average achieved speedup from 4.13x to 4.89x, while the average resource footprint (BRAM, DSP, FF and LUT) decreases from 93\% to 75\%. This result indicates that increasing backbone capacity provides a modest QoR benefit within the same MailoHLS pipeline. The improvement is not dramatic, which suggests that the gains of MailoHLS do not depend solely on scaling the base LLM, but arise from the robustness of the overall framework, namely the combination of semantic reasoning, structural fusion through cross-attention, and objective-aware adaptation.

\noindent\textbf{Cross-attention placement:} We further study the effect of structural-fusion density by varying the insertion interval of cross-attention layers within the LLM. For the 32-layer DeepSeek-Coder-7B backbone, inserting cross-attention every 8 layers corresponds to 4 fusion points, while every 16 layers corresponds to only 2. Denser fusion improves the average speedup from 3.58x to 4.89x, at the cost of a higher total resource footprint from 68.0\% to 93.3\%. More frequent cross-attention strengthens the influence of structural memory and sharpens objective-specific behavior, especially for aggressive latency-oriented designs. Conversely, with fewer insertion points, MailoHLS remains effective but behaves more conservatively, resembling the simpler SFT-only regime more closely than the fully structure-aligned model. In the remainder, we therefore use the every-8-layer configuration as the default setting.

\begin{figure}[t]
        \centering
        \includegraphics[width=\linewidth]{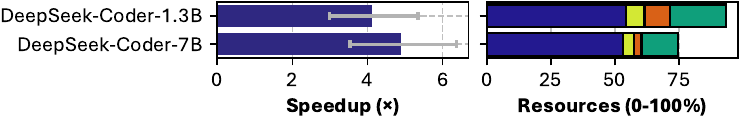}
        \caption{\framework' sensitivity to employed LLM backbone.}
    \label{fig:llm-backbone-impact-analysis}
\end{figure}

\subsubsection{Parameter-Efficient Objective-Aware Finetuning}
\label{subsubsec:peft}
\framework adopts a parameter-efficient finetuning (PEFT) approach.
Instead of updating all parameters, it employs LoRA~\cite{hu2022lora} to introduce a small number of trainable matrices, enabling efficient specialization while preserving the pretrained semantic capabilities of the backbone.
As shown in Figure~\ref{fig:llm-path}, LoRA modules are applied to both attention and feed-forward layers, allowing the model to focus on directive placeholders, objective tokens, and relevant code regions during directive prediction.

To support multiple optimization goals (e.g., latency, power, resource usage), \framework adopts a multi-adapter design, where each objective is handled by a dedicated LoRA adapter.
At runtime, the appropriate adapter is selected based on the target goal~\cite{vllm_lora_docs}.
Different objectives correspond to distinct trade-offs in the design space and often require conflicting directive configurations.
For instance, minimizing latency increases parallelism and resource usage, while optimizing for power or area favors more constrained configurations.
Thus, isolating adaptation per objective enables specialization across different regions of the Pareto frontier.

The training of each adapter is performed in three stages to decouple \textit{i)} semantic alignment, \textit{ii)} structural alignment, and \textit{iii)} objective-aware preference modeling.
In the first stage, supervised fine-tuning (SFT)~\cite{chen2024sft} is applied to the LoRA matrices, training the model to imitate directive assignments from labeled examples. 
Training data consists of directive configurations evaluated through HLS, where each sample includes an annotated kernel and a directive assignment corresponding to a valid design point. 
This stage enables the model to learn a semantic mapping from code and objectives to feasible pragma configurations.
In the second stage, LoRA adapters are frozen and SFT is applied only to the cross-attention layers, aligning structural embeddings from the GNN with the LLM representations. 
This stage enables the model to incorporate hardware-relevant constraints such as dependencies, parallelism, and memory behavior into its predictions without altering the learned semantic mapping.
In the final stage, Direct Preference Optimization (DPO)~\cite{rafailov2023dpo} is applied to the cross-attention layers, learning to prefer better configurations over worse ones under the objective associated with the active adapter.
Preference pairs are constructed from Pareto-optimal design points by selecting configurations that achieve better trade-offs under the target objective and contrasting them with dominated or inferior alternatives. This stage enables objective-aware learning, guiding the model toward directive choices closer to the desired region of the Pareto frontier.
\section{Evaluation Setup}

\subsection{Training Dataset}
For training and evaluation, we leverage the GN$\Omega$SIS dataset~\cite{ferikoglou2026gnomegasis}, one of the largest publicly available collections of HLS design points.
By applying a diverse set of HLS directives, it captures both latency and resource utilization, resulting in a rich design space.
However, GN$\Omega$SIS does not provide power measurements, and thus power is not considered as an optimization objective in this work.
Overall, it comprises nearly 219K design points across two FPGA architectures and three target clock frequencies.
We focus on the subset corresponding to the UltraScale+ ZCU104 platform at 100 MHz, which includes approximately 26,500 design points.

\subsection{Training Parameters}
We train one LoRA adapter per optimization objective using the three-stage pipeline described in \S\ref{subsubsec:peft}.
All adapters share a 4-bit quantized decoder-only backbone (NF4, bfloat16), namely Deepseek Coder 7B, with PagedAdamW8bit~\cite{dettmers20228bitoptimizers}, cosine scheduling, and 3\% warmup; gradient checkpointing supports 4K context prompts.

\noindent$\blacktriangleright$~\textbf{Stage 1 (Semantic Alignment):} We apply SFT on objective-conditioned directive completion, with cross-attention disabled.
LoRA ($r=8, \alpha=16$) is applied to the LLM and trained for 4 epochs with a learning rate of $5\times10^{-5}$.
We retain the top-6 design points per (kernel, objective) with score-based weighting.

\noindent$\blacktriangleright$~\textbf{Stage 2 (Structural Alignment):} Cross-attention is enabled to inject GNN embeddings, while LoRA and embeddings are frozen. 
Cross-attention is inserted every 8 layers (4 heads, dim 64), with structural memory of 64 slots (dim 32).
Only cross-attention and gating parameters are trained for 4 epochs (lr=$10^{-4}$ and $2\times10^{-4}$).

\noindent$\blacktriangleright$~\textbf{Stage 3 (DPO):} We apply DPO on Pareto-ranked preference pairs. 
LoRA remains frozen; only cross-attention and gating are updated (lr=$10^{-5}$ and $2\times10^{-5}$ for 3 epochs).

\noindent\framework is publicly available as an open-source project\footnote{Omitted for blind review}.

\subsection{Baselines and State-of-the-Art}
\label{subsec:6.3:baselinesSotA}

We evaluate \framework against a diverse set of baselines. 
As a reference, we consider the original source code without any HLS directives, as generated by VitisHLS. 
Furthermore, for each application in the GN$\Omega$SIS dataset~\cite{ferikoglou2026gnomegasis}, we extract the Pareto frontier and identify three representative design points: (a) the latency-optimal, (b) the resource-optimal, and (c) the Pareto knee, which captures a balanced trade-off between performance and resource utilization. 
These points serve as strong baselines for evaluating \framework across different optimization objectives.
We also compare \framework against CollectiveHLS~\cite{ferikoglou2024collectivehls}, an open SotA data-driven approach that leverages the GN$\Omega$SIS dataset to efficiently propose high-quality HLS directives. 
In addition, we benchmark against general-purpose large language models, including GPT-4o~\cite{openai2023gpt4}, Claude Haiku 4.5~\cite{anthropic2024claude3}, and Gemini 2.5 Pro~\cite{islam2024gemini}, which are prompted to generate directive configurations.
Finally, we include a comparison with LIFT~\cite{prakriya2025lift}, a SotA LLM-based framework for HLS optimization. LIFT integrates LLMs with GNN-based representations to capture both the sequential and structural characteristics of code, enabling the generation of performance-critical directives.
\section{Experimental Results}

\subsection{MailoHLS Evaluation}

\subsubsection{Evaluation on seen kernels}

\begin{figure*}[t]
    \centering
    \begin{subfigure}[t]{0.33\linewidth}
        \centering
    \includegraphics[width=\textwidth]{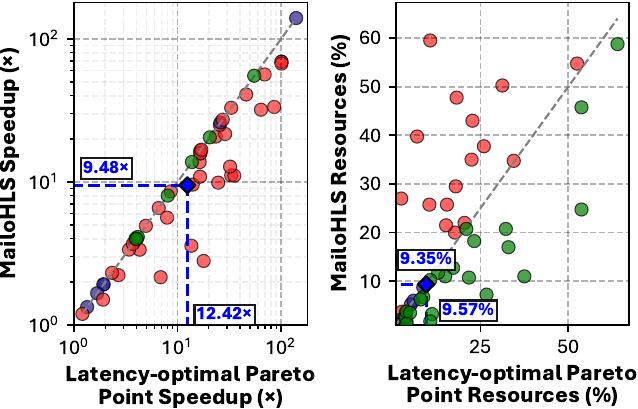}
    \caption{Latency-optimized Adapter}
    \end{subfigure}
    \begin{subfigure}[t]{.33\linewidth}
        \centering
    \includegraphics[width=\textwidth]{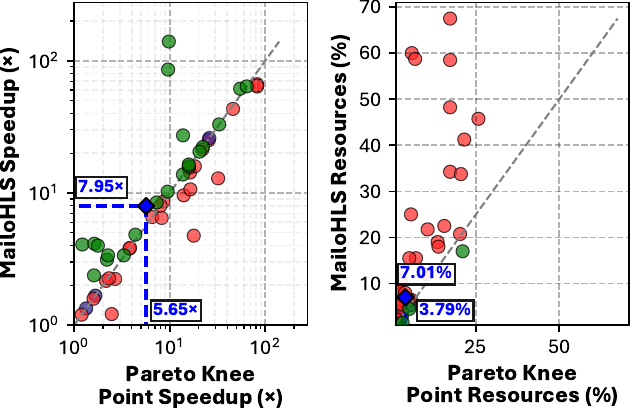}
    \caption{Balanced-optimized Adapter}
    \end{subfigure}
    \begin{subfigure}[t]{.33\linewidth}
        \centering
    \includegraphics[width=\textwidth]{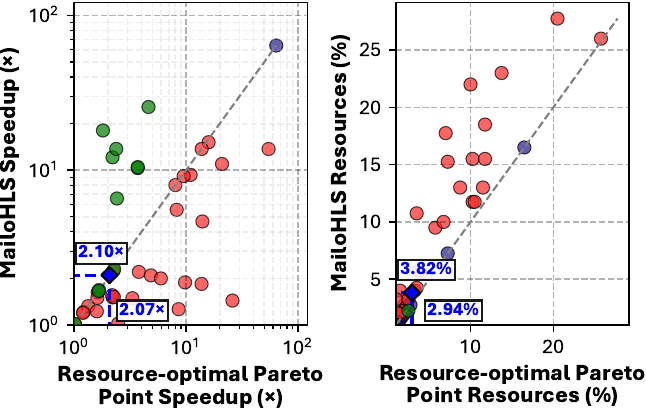}
    \caption{Resources-optimized Adapter}
    \end{subfigure}

    \caption{\framework's evaluation under an 80/10/10 dataset split, where designs generated by the latency-, balanced-, and resource-optimized adapters are compared against GN$\Omega$SIS Pareto reference points in terms of speedup and resource utilization.}
    \label{fig:mailos-eval-seen}
\end{figure*}

We partition the dataset into 80\%, 10\%, and 10\% splits for training, validation, and testing.
We train \framework to generate directive configurations conditioned on different optimization objectives, namely latency-optimized, resource-optimized, and balanced designs.
The generated designs are synthesized using Vitis HLS 2021.1 to obtain latency and resource utilization metrics, which are then compared against the corresponding Pareto reference points (latency-optimal, Pareto knee, and resource-optimal) derived directly from GN$\Omega$SIS dataset.
Figure~\ref{fig:mailos-eval-seen} shows, for each objective, the speedup over Vitis and resource utilization relative to the Pareto reference. Diagonal points indicate optimal designs, while deviations reflect latency–resource trade-offs. Green points denote cases where \framework dominates the reference (higher speedup and/or lower resources), highlighting the incompleteness of the GN$\Omega$SIS Pareto set, whereas red points indicate under-performance.

For the latency-optimized adapter, \framework achieves a geometric mean speedup of $9.48\times$ compared to $12.42\times$ for the corresponding Pareto reference (0.76$\times$ relative), while maintaining slightly lower average resource utilization (9.35\% vs. 9.75\%).
Most design points lie close to the diagonal, indicating that \framework consistently approximates the target region of the design space.
In several cases, the generated configurations dominate the reference points, demonstrating the ability of \framework to uncover high-quality designs beyond those identified by the GN$\Omega$SIS exploration.
For the balanced-optimized adapter, \framework attains a geometric mean speedup of $7.95\times$, exceeding the Pareto knee reference of $5.65\times$, at the cost of higher average resource utilization (7.01\% vs. 3.79\%).
Finally, for the resource-optimized adapter, \framework achieves a geometric mean speedup of $2.10\times$, comparable to the reference value of $2.07\times$, with a marginal increase in resource utilization (3.82\% vs. 2.94\%).
In this case, the distribution of points is more dispersed, due to the discrete and non-smooth nature of resource metrics in HLS, which makes precise control of utilization more challenging than latency-driven optimization.

To better understand the sources of design quality, we analyze the directives generated by \framework across the different optimization objectives. 
For the seen kernels, the latency-optimized adapter applies \texttt{\#pragma HLS PIPELINE II=1} to 34.8\% of loop structures, compared to 31.9\% for the balanced adapter and only 15.9\% for the resource-optimized adapter.
A similar trend is observed in loop unrolling. 
Aggressive unrolling factors are employed in 17.4\% of latency-oriented loops and 11.6\% of balanced loops, while they are not used in the resource-oriented setting. 
This behavior reflects the model’s tendency to exploit parallelism more aggressively when optimizing for performance.
The same pattern extends to memory optimizations. 
Large array partitioning factors are predicted for 7.1\% of arrays in the latency adapter, compared to 3.6\% and 1.8\% for the balanced and resource-optimized adapters, respectively.
In contrast, the choice of partitioning type (e.g., cyclic or block) remains relatively consistent across objectives. 
This suggests that objective-specific specialization is primarily achieved through varying the intensity of loop-level parallelism and memory banking, rather than through selecting different classes of directives.

\subsubsection{Evaluation on unseen kernels}

\begin{figure}[t]
    \centering
    \includegraphics[width=\linewidth]{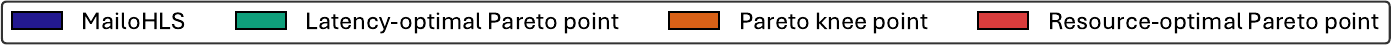}
    \begin{subfigure}[t]{\linewidth}
        \centering
    \includegraphics[width=\textwidth]{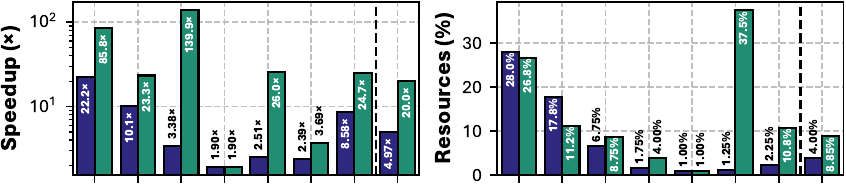}
    \caption{Latency-optimized Adapter}
    \end{subfigure}
    \begin{subfigure}[t]{\linewidth}
        \centering
    \includegraphics[width=\textwidth]{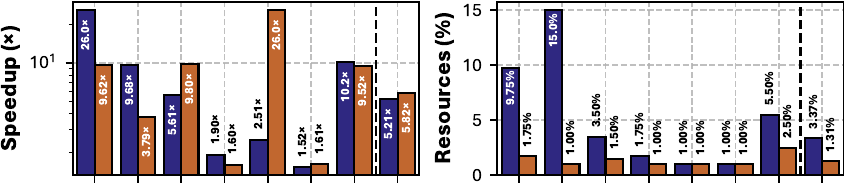}
    \caption{Balanced-optimized Adapter}
    \end{subfigure}
    \begin{subfigure}[t]{\linewidth}
        \centering
    \includegraphics[width=\textwidth]{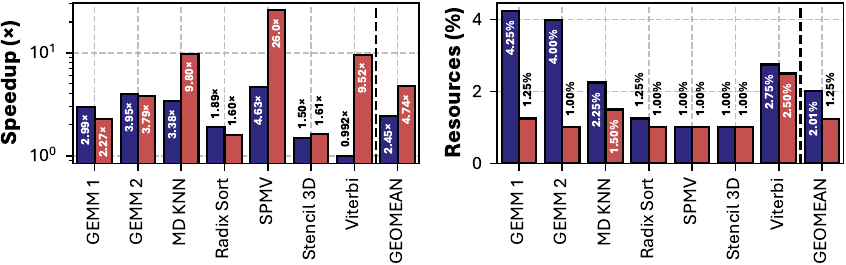}
    \caption{Resources-optimized Adapter}
    \end{subfigure}

    \caption{Speedup and Resources on Unseen Kernel Suite.}
    \label{fig:mailos-eval-unseen}
\end{figure}

To evaluate generalization, we adopt a leave-one-family-out setting in which the entire MachSuite kernels family is excluded from training and used solely for testing. 
Figure~\ref{fig:mailos-eval-unseen} reports the achieved speedup over Vitis alongside total resource utilization normalized to 0–100\% for the three adapters. 
The latency-optimized adapter achieves a geometric mean speedup of $4.97\times$, below the $20\times$ Pareto reference, while using significantly fewer resources (4\% vs. 8.85\%).
The balanced adapter achieves a geometric mean speedup of $5.21\times$, closely approaching the Pareto baseline of $5.82\times$, with an average resource utilization of 3.37\% compared to 1.31\%. 
As expected, it occupies the intermediate region of the latency–resource trade-off space. 
Notably, it often proves more robust, in some cases matching or surpassing the latency-optimized adapter while avoiding higher resource overhead.
The resource-optimized adapter achieves a geometric mean speedup of $2.45\times$ compared to $4.74\times$ for the Pareto baseline, with an average resource utilization of 2.01\% versus 1.25\%. 
While a performance gap remains, the results demonstrate that \framework retains effective control over resource usage even under distribution shift, albeit with reduced precision relative to seen kernels.

These results further reveal that the three adapters learn and transfer distinct optimization policies. 
The latency adapter favors aggressive pipelining, loop-level parallelism, and memory partitioning when exploitable concurrency is inferred. 
In contrast, the balanced adapter exhibits greater robustness by avoiding extreme transformations, achieving competitive performance while maintaining moderate resource usage. 
The resource-oriented adapter deliberately constrains hardware replication and memory pressure, resulting in lower area at the cost of underutilized parallelism in compute-bound kernels.
Finally, the larger gap to the Pareto reference in the unseen setting reflects the inherent difficulty of transferring hardware-aware optimization strategies (e.g., scheduling, parallelism, and memory access optimization). Nevertheless, \framework consistently outperforms the directive-free baseline and remains well aligned with the target objective. This indicates that \framework captures transferable micro-architectural principles, enabling effective generalization to unseen kernels without retraining or per-kernel exploration.

\vspace{-10pt}
\subsubsection{Ablation Study}

\begin{figure}[t]
    \centering
    \includegraphics[width=.6\linewidth]{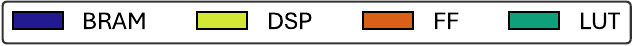}
    \includegraphics[width=\linewidth]{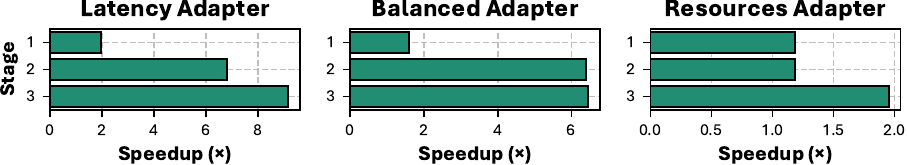}
    \includegraphics[width=\linewidth]{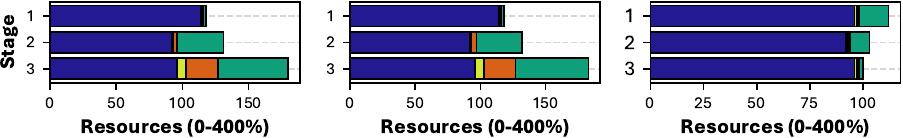}
    \caption{Impact of \framework' training stages.}
    \label{fig:mailos-eval-ablation}
\end{figure}

To isolate the contribution of each training stage, we conduct an ablation study on the Kalman kernel, one of the most structurally demanding applications in our evaluation due to its deeply nested loops, large on-chip buffers, and tight interplay between loop-level parallelism and memory partitioning.
Figure~\ref{fig:mailos-eval-ablation} illustrates the achieved speedup over Vitis alongside resource utilization across the different stages of the training pipeline for the three objective-specific adapters.

Objective-conditioned SFT (\textit{Step 1}) on the LoRA adapters produces only modest improvements, achieving speedups of $2\times$, $1.7\times$, and $1.2\times$ for the latency, balanced, and resource-oriented adapters, respectively. 
The latency and balanced adapters over-utilize BRAM, while only the resource-oriented adapter respects FPGA constraints, indicating that semantic code understanding alone is insufficient to fully capture the structural limits required for aggressive pipelining, unrolling, and memory partitioning.
The largest gains occur when GNN-derived structural information is incorporated through cross-attention (\textit{Step 2}). 
At this stage, the latency, balanced, and resource adapters achieve $6.3\times$, $6.2\times$, and $1.15\times$ speedups, respectively. Exposure to loop hierarchy, dependency structure, and memory interactions enables the model to apply far more effective micro-architectural transformations. 
Although these improvements come with a slight increase in hardware usage, both the latency and balanced adapters now produce BRAM-compliant designs, highlighting the critical role of structural awareness.
Finally, DPO (\textit{Step 3}) aligns the generated designs with the target objectives, achieving speedups of $8.6\times$, $6.2\times$, and $1.9\times$ for the latency-, balanced-, and resource-oriented adapters. 
For the latency adapter, DPO further boosts performance by steering designs toward higher-throughput configurations. 
The balanced adapter preserves most of the Stage 2 gains while controlling resource growth, consistent with its knee-region behavior. For the resource adapter, DPO enforces conservative resource usage while improving speedup relative to earlier stages. 
Overall, DPO effectively guides the final designs toward the intended Pareto-optimal trade-offs for each objective.

\subsection{Comparison with State-of-the-Art}

In the final stage of our analysis, we evaluate \framework against prior SotA HLS optimization methods as well as high-end LLMs (c.f.~\ref{subsec:6.3:baselinesSotA}). 
We use as driver applications the Kalman Filter and GAN~\cite{danopoulos2021fpga}, which stress complementary aspects of HLS optimization. 
The Kalman Filter kernel is memory-bound, with performance primarily limited by data movement, memory access patterns, and on-chip storage efficiency, whereas the GAN kernel is compute-bound, driven by arithmetic intensity, parallel execution of operations, and the ability to fully utilize available compute resources.

\begin{figure}[t]
    \centering
    \includegraphics[width=.6\linewidth]{figures/04_Experimental/mailos-eval-ablation-legend-crop.pdf}
    \\
    \begin{subfigure}[t]{\linewidth}
    \centering
    \includegraphics[width=\linewidth]{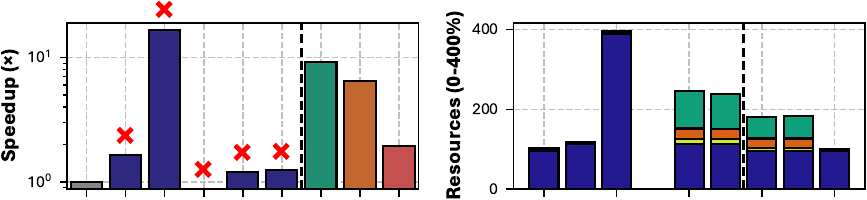}
    \caption{Kalman Filter}
    \label{fig:mailos-eval-sota-kalman}
    \end{subfigure}
    \begin{subfigure}[t]{\linewidth}
    \centering
    \includegraphics[width=\linewidth]{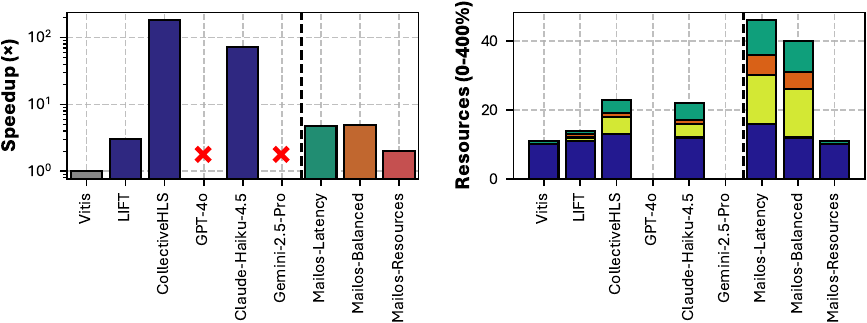}
    \caption{GAN}
    \label{fig:mailos-eval-sota-gan}
    \end{subfigure}

    \caption{Comparison with state-of-the-art on a memory- (Kalman) and a compute-bound (GAN) application.}
    \label{fig:mailos-eval-sota}
\end{figure}

Figure~\ref{fig:mailos-eval-sota} shows the respective results.
For Kalman, \framework delivers the strongest combination of design validity, quality of results, and alignment with optimization objectives. 
It produces valid designs achieving speedups of $8.6\times$, $6.2\times$, and $1.9\times$ for the latency-, balanced-, and resource-optimized targets, respectively, all while respecting the FPGA’s resource limits. 
In contrast, general-purpose LLMs prove unreliable: GPT-4o fails to generate any design, while Claude Haiku 4.5 and Gemini 2.5 Pro produce configurations that exceed FPGA capacity, reflecting their inability to handle the complexity of memory-bound applications. 
CollectiveHLS achieves high speedups on Kalman but requires excessive resources, rendering the designs non-feasible.
LIFT, though structurally aware and closer to feasible designs, still over-utilizes BRAM due to incomplete handling of array partitioning directives.
For the GAN application, which is MAC-intensive, highly aggressive loop parallelism can generate large speedups if activation buffers are sufficiently partitioned to sustain bandwidth. 
CollectiveHLS and Claude Haiku 4.5 achieve remarkable speedups of $186.8\times$ and $72.9\times$, respectively, while remaining within FPGA resource limits, showing that GAN favors extreme throughput-oriented transformations. 
While \framework does not reach these extremes, it is the only method to produce three explicit objective-conditioned operating points from a single framework. 
It achieves $4.7\times$, $4.9\times$, and $2\times$ speedups for the latency-, balanced-, and resource-oriented targets, all within the FPGA’s capacity.
The performance gap with CollectiveHLS and Claude Haiku 4.5 can be attributed to differing optimization strategies.
In this case, \framework favors explicit loop unrolling and array partitioning, whereas the competing methods rely more heavily on pipelining. 
In particular, applying pipelining at outer loops can implicitly trigger full unrolling of inner loops, significantly increasing parallelism and yielding higher speedups.
Such aggressive transformations are not adopted by \framework, as they  risk violating resource constraints and leading to non-feasible designs. 
Instead, \framework adopts a more conservative strategy, guided by feasibility considerations learned from prior design points.
Last, \framework outperforms LIFT, which achieves only a $3.1\times$ speedup, particularly for the latency- and balanced-oriented designs, demonstrating its ability to generate efficient, well-balanced configurations across objectives.

\section{Conclusion}

We present \framework, a unified approach for objective-aware HLS directive optimization that combines semantic reasoning with GNN-based structural modeling. By integrating structural embeddings via cross-attention and leveraging PEFT with objective-conditioned LoRA adapters and Pareto-driven optimization, \framework enables joint reasoning over code semantics, structure, and design trade-offs. Across seen and unseen kernels, it achieves up to $12.42\times$ and $8.4\times$ speedup ($9.48\times$ and $4.97\times$ geometric mean), consistently producing near-Pareto-optimal designs. On unseen applications, it reaches up to $10.2\times$ ($6.58\times$ geometric mean), outperforming prior methods and general-purpose LLMs while demonstrating strong generalization.



\bibliographystyle{ACM-Reference-Format}
\bibliography{refs-base,refs-gnn-papers,refs-llm-papers,refs-synthesis-based-papers}

@inproceedings{tomkou2026linking,
  title={Linking High-Level Synthesis with FPGA Runtime Orchestration},
  author={Tomkou, Despoina and Ferikoglou, Aggelos and Masouros, Dimosthenis and Xydis, Sotirios and Soudris, Dimitrios},
  booktitle={17th Workshop on Parallel Programming and Run-Time Management Techniques for Many-Core Architectures and 15th Workshop on Design Tools and Architectures for Multicore Embedded Computing Platforms (PARMA-DITAM 2026)},
  pages={7--1},
  year={2026},
  organization={Schloss Dagstuhl--Leibniz-Zentrum f{\"u}r Informatik}
}

@inproceedings{fu2023gpt4aigchip,
  title={Gpt4aigchip: Towards next-generation ai accelerator design automation via large language models},
  author={Fu, Yonggan and Zhang, Yongan and Yu, Zhongzhi and Li, Sixu and Ye, Zhifan and Li, Chaojian and Wan, Cheng and Lin, Yingyan Celine},
  booktitle={2023 IEEE/ACM International Conference on Computer Aided Design (ICCAD)},
  pages={1--9},
  year={2023},
  organization={IEEE}
}

@article{wang2025llm,
  title={LLM-DSE: Searching Accelerator Parameters with LLM Agents},
  author={Wang, Hanyu and Wu, Xinrui and Ding, Zijian and Zheng, Su and Wang, Chengyue and Prakriya, Neha and Nowatzki, Tony and Sun, Yizhou and Cong, Jason},
  journal={arXiv preprint arXiv:2505.12188},
  year={2025}
}

@inproceedings{rafailov2023dpo,
  author       = {Rafael Rafailov and
                  Archit Sharma and
                  Eric Mitchell and
                  Christopher D. Manning and
                  Stefano Ermon and
                  Chelsea Finn},
  editor       = {Alice Oh and
                  Tristan Naumann and
                  Amir Globerson and
                  Kate Saenko and
                  Moritz Hardt and
                  Sergey Levine},
  title        = {Direct Preference Optimization: Your Language Model is Secretly a
                  Reward Model},
  booktitle    = {Advances in Neural Information Processing Systems 36: Annual Conference
                  on Neural Information Processing Systems 2023, NeurIPS 2023, New Orleans,
                  LA, USA, December 10 - 16, 2023},
  year         = {2023},
  url          = {http://papers.nips.cc/paper\_files/paper/2023/hash/a85b405ed65c6477a4fe8302b5e06ce7-Abstract-Conference.html},
  bibsource    = {dblp computer science bibliography, https://dblp.org}
}

@inproceedings{chen2024sft,
  author       = {Zixiang Chen and
                  Yihe Deng and
                  Huizhuo Yuan and
                  Kaixuan Ji and
                  Quanquan Gu},
  editor       = {Ruslan Salakhutdinov and
                  Zico Kolter and
                  Katherine A. Heller and
                  Adrian Weller and
                  Nuria Oliver and
                  Jonathan Scarlett and
                  Felix Berkenkamp},
  title        = {Self-Play Fine-Tuning Converts Weak Language Models to Strong Language
                  Models},
  booktitle    = {Forty-first International Conference on Machine Learning, {ICML} 2024,
                  Vienna, Austria, July 21-27, 2024},
  series       = {Proceedings of Machine Learning Research},
  volume       = {235},
  pages        = {6621--6642},
  publisher    = {{PMLR} / OpenReview.net},
  year         = {2024},
  url          = {https://proceedings.mlr.press/v235/chen24j.html},
  bibsource    = {dblp computer science bibliography, https://dblp.org}
}

@inproceedings{lattner2004llvm,
  author       = {Chris Lattner and
                  Vikram S. Adve},
  title        = {{LLVM:} {A} Compilation Framework for Lifelong Program Analysis {\&}
                  Transformation},
  booktitle    = {2nd {IEEE} / {ACM} International Symposium on Code Generation and
                  Optimization {(CGO} 2004), 20-24 March 2004, San Jose, CA, {USA}},
  pages        = {75--88},
  publisher    = {{IEEE} Computer Society},
  year         = {2004},
  url          = {https://doi.org/10.1109/CGO.2004.1281665},
  doi          = {10.1109/CGO.2004.1281665},
  bibsource    = {dblp computer science bibliography, https://dblp.org}
}

@inproceedings{hu2022lora,
  author       = {Edward J. Hu and
                  Yelong Shen and
                  Phillip Wallis and
                  Zeyuan Allen{-}Zhu and
                  Yuanzhi Li and
                  Shean Wang and
                  Lu Wang and
                  Weizhu Chen},
  title        = {LoRA: Low-Rank Adaptation of Large Language Models},
  booktitle    = {The Tenth International Conference on Learning Representations, {ICLR}
                  2022, Virtual Event, April 25-29, 2022},
  publisher    = {OpenReview.net},
  year         = {2022},
  url          = {https://openreview.net/forum?id=nZeVKeeFYf9},
  bibsource    = {dblp computer science bibliography, https://dblp.org}
}

@inproceedings{vaswani2017attention,
  author       = {Ashish Vaswani and
                  Noam Shazeer and
                  Niki Parmar and
                  Jakob Uszkoreit and
                  Llion Jones and
                  Aidan N. Gomez and
                  Lukasz Kaiser and
                  Illia Polosukhin},
  editor       = {Isabelle Guyon and
                  Ulrike von Luxburg and
                  Samy Bengio and
                  Hanna M. Wallach and
                  Rob Fergus and
                  S. V. N. Vishwanathan and
                  Roman Garnett},
  title        = {Attention is All you Need},
  booktitle    = {Advances in Neural Information Processing Systems 30: Annual Conference
                  on Neural Information Processing Systems 2017, December 4-9, 2017,
                  Long Beach, CA, {USA}},
  pages        = {5998--6008},
  year         = {2017},
  url          = {https://proceedings.neurips.cc/paper/2017/hash/3f5ee243547dee91fbd053c1c4a845aa-Abstract.html},
  bibsource    = {dblp computer science bibliography, https://dblp.org}
}

@inproceedings{basalama2025streamhls,
  author       = {Suhail Basalama and
                  Jason Cong},
  editor       = {Andrew Putnam and
                  Jing Li},
  title        = {Stream-HLS: Towards Automatic Dataflow Acceleration},
  booktitle    = {Proceedings of the 2025 {ACM/SIGDA} International Symposium on Field
                  Programmable Gate Arrays, {FPGA} 2025, Monterey, CA, USA, 27 February
                  2025 - 1 March 2025},
  pages        = {103--114},
  publisher    = {{ACM}},
  year         = {2025},
  url          = {https://doi.org/10.1145/3706628.3708878},
  doi          = {10.1145/3706628.3708878},
  bibsource    = {dblp computer science bibliography, https://dblp.org}
}

@article{putnam2014reconfigurable,
  title={A reconfigurable fabric for accelerating large-scale datacenter services},
  author={Putnam, Andrew and Caulfield, Adrian M and Chung, Eric S and Chiou, Derek and Constantinides, Kypros and Demme, John and Esmaeilzadeh, Hadi and Fowers, Jeremy and Gopal, Gopi Prashanth and Gray, Jan and others},
  journal={ACM SIGARCH Computer Architecture News},
  volume={42},
  number={3},
  pages={13--24},
  year={2014},
  publisher={ACM New York, NY, USA}
}

@article{bobda2022future,
  title={The future of FPGA acceleration in datacenters and the cloud},
  author={Bobda, Christophe and Mbongue, Joel Mandebi and Chow, Paul and Ewais, Mohammad and Tarafdar, Naif and Vega, Juan Camilo and Eguro, Ken and Koch, Dirk and Handagala, Suranga and Leeser, Miriam and others},
  journal={ACM Transactions on Reconfigurable Technology and Systems (TRETS)},
  volume={15},
  number={3},
  pages={1--42},
  year={2022},
  publisher={ACM New York, NY}
}

@misc{amazon_f1_instance,
  author       = {{Amazon Web Services}},
  title        = {EC2 Instances (F1) with Programmable Hardware},
  year         = {2016},
  howpublished = {\url{https://aws.amazon.com/blogs/aws/developer-preview-ec2-instances-f1-with-programmable-hardware/}},
  note         = {Accessed: 2025-10-25}
}

@article{wan2021survey,
  title={A survey of fpga-based robotic computing},
  author={Wan, Zishen and Yu, Bo and Li, Thomas Yuang and Tang, Jie and Zhu, Yuhao and Wang, Yu and Raychowdhury, Arijit and Liu, Shaoshan},
  journal={IEEE Circuits and Systems Magazine},
  volume={21},
  number={2},
  pages={48--74},
  year={2021},
  publisher={IEEE}
}

@article{chamola2020fpga,
  title={FPGA for 5G: Re-configurable hardware for next generation communication},
  author={Chamola, Vinay and Patra, Sambit and Kumar, Neeraj and Guizani, Mohsen},
  journal={IEEE Wireless Communications},
  volume={27},
  number={3},
  pages={140--147},
  year={2020},
  publisher={IEEE}
}

@article{xydis2014spirit,
  title={SPIRIT: Spectral-aware Pareto iterative refinement optimization for supervised high-level synthesis},
  author={Xydis, Sotirios and Palermo, Gianluca and Zaccaria, Vittorio and Silvano, Cristina},
  journal={IEEE Transactions on Computer-Aided Design of Integrated Circuits and Systems},
  volume={34},
  number={1},
  pages={155--159},
  year={2014},
  publisher={IEEE}
}

@inproceedings{dai2018fast,
  title={Fast and accurate estimation of quality of results in high-level synthesis with machine learning},
  author={Dai, Steve and Zhou, Yuan and Zhang, Hang and Ustun, Ecenur and Young, Evangeline FY and Zhang, Zhiru},
  booktitle={2018 IEEE 26th Annual International Symposium on Field-Programmable Custom Computing Machines (FCCM)},
  pages={129--132},
  year={2018},
  organization={IEEE}
}

@inproceedings{makrani2019xppe,
  title={Xppe: cross-platform performance estimation of hardware accelerators using machine learning},
  author={Makrani, Hosein Mohammadi and Sayadi, Hossein and Mohsenin, Tinoosh and Rafatirad, Setareh and Sasan, Avesta and Homayoun, Houman},
  booktitle={Proceedings of the 24th Asia and South Pacific Design Automation Conference},
  pages={727--732},
  year={2019}
}

@inproceedings{lin2020hl,
  title={HL-Pow: A learning-based power modeling framework for high-level synthesis},
  author={Lin, Zhe and Zhao, Jieru and Sinha, Sharad and Zhang, Wei},
  booktitle={2020 25th Asia and South Pacific Design Automation Conference (ASP-DAC)},
  pages={574--580},
  year={2020},
  organization={IEEE}
}

@inproceedings{ustun2020accurate,
  title={Accurate operation delay prediction for FPGA HLS using graph neural networks},
  author={Ustun, Ecenur and Deng, Chenhui and Pal, Debjit and Li, Zhijing and Zhang, Zhiru},
  booktitle={Proceedings of the 39th international conference on computer-aided design},
  pages={1--9},
  year={2020}
}

@article{ferikoglou2023collectivehls,
  title={CollectiveHLS: Ultrafast Knowledge-Based HLS Design Optimization},
  author={Ferikoglou, Aggelos and Kakolyris, Andreas and Kypriotis, Vasilis and Masouros, Dimosthenis and Soudris, Dimitrios and Xydis, Sotirios},
  journal={IEEE Embedded Systems Letters},
  volume={16},
  number={2},
  pages={235--238},
  year={2023},
  publisher={IEEE}
}

@article{ferretti2021db4hls,
  title={Db4hls: A database of high-level synthesis design space explorations},
  author={Ferretti, Lorenzo and Kwon, Jihye and Ansaloni, Giovanni and Di Guglielmo, Giuseppe and Carloni, Luca and Pozzi, Laura},
  journal={IEEE Embedded Systems Letters},
  volume={13},
  number={4},
  pages={194--197},
  year={2021},
  publisher={IEEE}
}

@article{cong2011high,
  title={High-level synthesis for FPGAs: From prototyping to deployment},
  author={Cong, Jason and Liu, Bin and Neuendorffer, Stephen and Noguera, Juanjo and Vissers, Kees and Zhang, Zhiru},
  journal={IEEE Transactions on Computer-Aided Design of Integrated Circuits and Systems},
  volume={30},
  number={4},
  pages={473--491},
  year={2011},
  publisher={IEEE}
}

@inproceedings{papakonstantinou2011multilevel,
  title={Multilevel granularity parallelism synthesis on FPGAs},
  author={Papakonstantinou, Alexandros and Liang, Yun and Stratton, John A and Gururaj, Karthik and Chen, Deming and Hwu, Wen-Mei W and Cong, Jason},
  booktitle={2011 IEEE 19th Annual International Symposium on Field-Programmable Custom Computing Machines},
  pages={178--185},
  year={2011},
  organization={IEEE}
}

@inproceedings{rupnow2011study,
  title={A study of high-level synthesis: Promises and challenges},
  author={Rupnow, Kyle and Liang, Yun and Li, Yinan and Chen, Deming},
  booktitle={2011 9th IEEE International Conference on ASIC},
  pages={1102--1105},
  year={2011},
  organization={IEEE}
}

@article{chi2022democratizing,
  title={Democratizing domain-specific computing},
  author={Chi, Yuze and Qiao, Weikang and Sohrabizadeh, Atefeh and Wang, Jie and Cong, Jason},
  journal={Communications of the ACM},
  volume={66},
  number={1},
  pages={74--85},
  year={2022},
  publisher={ACM New York, NY, USA}
}

@article{numan2020towards,
  title={Towards automatic high-level code deployment on reconfigurable platforms: A survey of high-level synthesis tools and toolchains},
  author={Numan, Mostafa W and Phillips, Braden J and Puddy, Gavin S and Falkner, Katrina},
  journal={IEEE Access},
  volume={8},
  pages={174692--174722},
  year={2020},
  publisher={IEEE}
}

@article{schafer2019high,
  title={High-level synthesis design space exploration: Past, present, and future},
  author={Schafer, Benjamin Carrion and Wang, Zi},
  journal={IEEE Transactions on Computer-Aided Design of Integrated Circuits and Systems},
  volume={39},
  number={10},
  pages={2628--2639},
  year={2019},
  publisher={IEEE}
}

@article{guo2024deepseekcoder,
  author       = {Daya Guo and
                  Qihao Zhu and
                  Dejian Yang and
                  Zhenda Xie and
                  Kai Dong and
                  Wentao Zhang and
                  Guanting Chen and
                  Xiao Bi and
                  Y. Wu and
                  Y. K. Li and
                  Fuli Luo and
                  Yingfei Xiong and
                  Wenfeng Liang},
  title        = {DeepSeek-Coder: When the Large Language Model Meets Programming -
                  The Rise of Code Intelligence},
  journal      = {CoRR},
  volume       = {abs/2401.14196},
  year         = {2024},
  url          = {https://doi.org/10.48550/arXiv.2401.14196},
  doi          = {10.48550/ARXIV.2401.14196},
  eprinttype   = {arXiv},
  eprint       = {2401.14196},
  bibsource    = {dblp computer science bibliography, https://dblp.org}
}

@article{ferikoglou2024collectivehls,
  title={CollectiveHLS: A collaborative approach to high-level synthesis design optimization},
  author={Ferikoglou, Aggelos and Kakolyris, Andreas and Masouros, Dimosthenis and Soudris, Dimitrios and Xydis, Sotirios},
  journal={ACM Transactions on Reconfigurable Technology and Systems},
  volume={18},
  number={1},
  pages={1--32},
  year={2024},
  publisher={ACM New York, NY}
}

@inproceedings{ferikoglou2024data,
  title={Data-driven HLS optimization for reconfigurable accelerators},
  author={Ferikoglou, Aggelos and Kakolyris, Andreas and Kypriotis, Vasilis and Masouros, Dimosthenis and Soudris, Dimitrios and Xydis, Sotirios},
  booktitle={Proceedings of the 61st ACM/IEEE Design Automation Conference},
  pages={1--6},
  year={2024}
}

@article{bai2023towards,
  title={Towards a comprehensive benchmark for high-level synthesis targeted to fpgas},
  author={Bai, Yunsheng and Sohrabizadeh, Atefeh and Qin, Zongyue and Hu, Ziniu and Sun, Yizhou and Cong, Jason},
  journal={Advances in Neural Information Processing Systems},
  volume={36},
  pages={45288--45299},
  year={2023}
}

@article{chang2024survey,
  title={A survey on evaluation of large language models},
  author={Chang, Yupeng and Wang, Xu and Wang, Jindong and Wu, Yuan and Yang, Linyi and Zhu, Kaijie and Chen, Hao and Yi, Xiaoyuan and Wang, Cunxiang and Wang, Yidong and others},
  journal={ACM transactions on intelligent systems and technology},
  volume={15},
  number={3},
  pages={1--45},
  year={2024},
  publisher={ACM New York, NY}
}

@inproceedings{cong2018understanding,
  title={Understanding performance differences of FPGAs and GPUs},
  author={Cong, Jason and Fang, Zhenman and Lo, Michael and Wang, Hanrui and Xu, Jingxian and Zhang, Shaochong},
  booktitle={2018 IEEE 26th annual international symposium on field-programmable custom computing machines (FCCM)},
  pages={93--96},
  year={2018},
  organization={IEEE}
}

@article{ferikoglou2026gnomegasis,
  title={GN$\Omega$SIS: Lessons Learned in Generating a High-Level Synthesis Dataset},
  author={Ferikoglou, Aggelos and Tomkou, Despoina and Masouros, Dimosthenis and Soudris, Dimitrios and Xydis, Sotirios},
  journal={ACM Transactions on Architecture and Code Optimization},
  year={2026},
  publisher={ACM New York, NY}
}

@article{keshar2019ctrl,
  author       = {Nitish Shirish Keskar and
                  Bryan McCann and
                  Lav R. Varshney and
                  Caiming Xiong and
                  Richard Socher},
  title        = {{CTRL:} {A} Conditional Transformer Language Model for Controllable
                  Generation},
  journal      = {CoRR},
  volume       = {abs/1909.05858},
  year         = {2019},
  url          = {http://arxiv.org/abs/1909.05858},
  eprinttype   = {arXiv},
  eprint       = {1909.05858},
  bibsource    = {dblp computer science bibliography, https://dblp.org}
}

@article{raffel2019t5,
  author       = {Colin Raffel and
                  Noam Shazeer and
                  Adam Roberts and
                  Katherine Lee and
                  Sharan Narang and
                  Michael Matena and
                  Yanqi Zhou and
                  Wei Li and
                  Peter J. Liu},
  title        = {Exploring the Limits of Transfer Learning with a Unified Text-to-Text
                  Transformer},
  journal      = {CoRR},
  volume       = {abs/1910.10683},
  year         = {2019},
  url          = {http://arxiv.org/abs/1910.10683},
  eprinttype   = {arXiv},
  eprint       = {1910.10683},
  bibsource    = {dblp computer science bibliography, https://dblp.org}
}

@inproceedings{wang2021codet5,
  author       = {Yue Wang and
                  Weishi Wang and
                  Shafiq R. Joty and
                  Steven C. H. Hoi},
  editor       = {Marie{-}Francine Moens and
                  Xuanjing Huang and
                  Lucia Specia and
                  Scott Wen{-}tau Yih},
  title        = {CodeT5: Identifier-aware Unified Pre-trained Encoder-Decoder Models
                  for Code Understanding and Generation},
  booktitle    = {Proceedings of the 2021 Conference on Empirical Methods in Natural
                  Language Processing, {EMNLP} 2021, Virtual Event / Punta Cana, Dominican
                  Republic, 7-11 November, 2021},
  pages        = {8696--8708},
  publisher    = {Association for Computational Linguistics},
  year         = {2021},
  url          = {https://doi.org/10.18653/v1/2021.emnlp-main.685},
  doi          = {10.18653/V1/2021.EMNLP-MAIN.685},
  bibsource    = {dblp computer science bibliography, https://dblp.org}
}

@article{cong2011hls,
  author       = {Jason Cong and
                  Bin Liu and
                  Stephen Neuendorffer and
                  Juanjo Noguera and
                  Kees A. Vissers and
                  Zhiru Zhang},
  title        = {High-Level Synthesis for FPGAs: From Prototyping to Deployment},
  journal      = {{IEEE} Trans. Comput. Aided Des. Integr. Circuits Syst.},
  volume       = {30},
  number       = {4},
  pages        = {473--491},
  year         = {2011},
  url          = {https://doi.org/10.1109/TCAD.2011.2110592},
  doi          = {10.1109/TCAD.2011.2110592},
  bibsource    = {dblp computer science bibliography, https://dblp.org}
}

@article{pan2025survey,
  title={A survey of research in large language models for electronic design automation},
  author={Pan, Jingyu and Zhou, Guanglei and Chang, Chen-Chia and Jacobson, Isaac and Hu, Jiang and Chen, Yiran},
  journal={ACM Transactions on Design Automation of Electronic Systems},
  volume={30},
  number={3},
  pages={1--21},
  year={2025},
  publisher={ACM New York, NY}
}

@article{ferretti2020leveraging,
  title={Leveraging prior knowledge for effective design-space exploration in high-level synthesis},
  author={Ferretti, Lorenzo and Kwon, Jihye and Ansaloni, Giovanni and Di Guglielmo, Giuseppe and Carloni, Luca P and Pozzi, Laura},
  journal={IEEE Transactions on Computer-Aided Design of Integrated Circuits and Systems},
  volume={39},
  number={11},
  pages={3736--3747},
  year={2020},
  publisher={IEEE}
}

@article{openai2023gpt4,
  author       = {OpenAI},
  title        = {{GPT-4} Technical Report},
  journal      = {CoRR},
  volume       = {abs/2303.08774},
  year         = {2023},
  url          = {https://doi.org/10.48550/arXiv.2303.08774},
  doi          = {10.48550/ARXIV.2303.08774},
  eprinttype   = {arXiv},
  eprint       = {2303.08774},
  bibsource    = {dblp computer science bibliography, https://dblp.org}
}

@misc{anthropic2024claude3,
  author       = {{Anthropic}},
  title        = {Model Card and Evaluations for Claude 3},
  year         = {2024},
  howpublished = {\url{https://www-cdn.anthropic.com/de8ba9b01c9ab7cbabf5c33b80b7bbc618857627/Model_Card_Claude_3.pdf}},
  note         = {Accessed: 2026-04-03}
}

@inproceedings{ye2022scalehls,
  title={Scalehls: A new scalable high-level synthesis framework on multi-level intermediate representation},
  author={Ye, Hanchen and Hao, Cong and Cheng, Jianyi and Jeong, Hyunmin and Huang, Jack and Neuendorffer, Stephen and Chen, Deming},
  booktitle={2022 IEEE international symposium on high-performance computer architecture (HPCA)},
  pages={741--755},
  year={2022},
  organization={IEEE}
}

@manual{vllm_lora_docs,
  title        = {vLLM Documentation: Dynamically Serving LoRA Adapters},
  author       = {{vLLM Team}},
  year         = {2024},
  url          = {https://docs.vllm.ai/en/stable/features/lora/#dynamically-serving-lora-adapters},
  note         = {Accessed: 2024-05-22}
}

@inproceedings{cummins2021programl,
  author       = {Chris Cummins and
                  Zacharias V. Fisches and
                  Tal Ben{-}Nun and
                  Torsten Hoefler and
                  Michael F. P. O'Boyle and
                  Hugh Leather},
  editor       = {Marina Meila and
                  Tong Zhang},
  title        = {ProGraML: {A} Graph-based Program Representation for Data Flow Analysis
                  and Compiler Optimizations},
  booktitle    = {Proceedings of the 38th International Conference on Machine Learning,
                  {ICML} 2021, 18-24 July 2021, Virtual Event},
  series       = {Proceedings of Machine Learning Research},
  pages        = {2244--2253},
  publisher    = {{PMLR}},
  year         = {2021},
  url          = {http://proceedings.mlr.press/v139/cummins21a.html},
  bibsource    = {dblp computer science bibliography, https://dblp.org}
}

@inproceedings{dettmers20228bitoptimizers,
  author       = {Tim Dettmers and
                  Mike Lewis and
                  Sam Shleifer and
                  Luke Zettlemoyer},
  title        = {8-bit Optimizers via Block-wise Quantization},
  booktitle    = {The Tenth International Conference on Learning Representations, {ICLR}
                  2022, Virtual Event, April 25-29, 2022},
  publisher    = {OpenReview.net},
  year         = {2022},
  url          = {https://openreview.net/forum?id=shpkpVXzo3h},
  bibsource    = {dblp computer science bibliography, https://dblp.org}
}

@article{lewis2020retrieval,
  title={Retrieval-augmented generation for knowledge-intensive nlp tasks},
  author={Lewis, Patrick and Perez, Ethan and Piktus, Aleksandra and Petroni, Fabio and Karpukhin, Vladimir and Goyal, Naman and K{\"u}ttler, Heinrich and Lewis, Mike and Yih, Wen-tau and Rockt{\"a}schel, Tim and others},
  journal={Advances in neural information processing systems},
  volume={33},
  pages={9459--9474},
  year={2020}
}

@inproceedings{danopoulos2021fpga,
  title={FPGA Acceleration of Generative Adversarial Networks for Image Reconstruction},
  author={Danopoulos, Dimitrios and Anagnostopoulos, Konstantinos and Kachris, Christoforos and Soudris, Dimitrios},
  booktitle={2021 10th International Conference on Modern Circuits and Systems Technologies (MOCAST)},
  pages={1--5},
  year={2021},
  organization={IEEE}
}

@inproceedings{xu2018jumping,
  author       = {Keyulu Xu and
                  Chengtao Li and
                  Yonglong Tian and
                  Tomohiro Sonobe and
                  Ken{-}ichi Kawarabayashi and
                  Stefanie Jegelka},
  editor       = {Jennifer G. Dy and
                  Andreas Krause},
  title        = {Representation Learning on Graphs with Jumping Knowledge Networks},
  booktitle    = {Proceedings of the 35th International Conference on Machine Learning,
                  {ICML} 2018, Stockholmsm{\"{a}}ssan, Stockholm, Sweden, July
                  10-15, 2018},
  series       = {Proceedings of Machine Learning Research},
  pages        = {5449--5458},
  publisher    = {{PMLR}},
  year         = {2018},
  url          = {http://proceedings.mlr.press/v80/xu18c.html},
  bibsource    = {dblp computer science bibliography, https://dblp.org}
}

@inproceedings{gao2024hierarchical,
  author       = {Mingzhe Gao and
                  Jieru Zhao and
                  Zhe Lin and
                  Minyi Guo},
  title        = {Hierarchical Source-to-Post-Route QoR Prediction in High-Level Synthesis
                  with GNNs},
  booktitle    = {Design, Automation {\&} Test in Europe Conference {\&} Exhibition,
                  {DATE} 2024, Valencia, Spain, March 25-27, 2024},
  pages        = {1--6},
  publisher    = {{IEEE}},
  year         = {2024},
  url          = {https://doi.org/10.23919/DATE58400.2024.10546555},
  doi          = {10.23919/DATE58400.2024.10546555},
  bibsource    = {dblp computer science bibliography, https://dblp.org}
}

@inproceedings{bai2024learning,
  author       = {Yunsheng Bai and
                  Atefeh Sohrabizadeh and
                  Zijian Ding and
                  Rongjian Liang and
                  Weikai Li and
                  Ding Wang and
                  Haoxing Ren and
                  Yizhou Sun and
                  Jason Cong},
  editor       = {Hussam Amrouch and
                  Jiang Hu and
                  Siddharth Garg and
                  Yibo Lin},
  title        = {Learning to Compare Hardware Designs for High-Level Synthesis},
  booktitle    = {Proceedings of the 2024 {ACM/IEEE} International Symposium on Machine
                  Learning for CAD, {MLCAD} 2024, Salt Lake City, UT, USA, September
                  9-11, 2024},
  pages        = {2},
  publisher    = {{ACM}},
  year         = {2024},
  url          = {https://doi.org/10.1145/3670474.3685940},
  doi          = {10.1145/3670474.3685940},
  bibsource    = {dblp computer science bibliography, https://dblp.org}
}

@inproceedings{li2025hierarchical,
  author       = {Weikai Li and
                  Ding Wang and
                  Zijian Ding and
                  Atefeh Sohrabizadeh and
                  Zongyue Qin and
                  Jason Cong and
                  Yizhou Sun},
  editor       = {Toby Walsh and
                  Julie Shah and
                  Zico Kolter},
  title        = {Hierarchical Mixture of Experts: Generalizable Learning for High-Level
                  Synthesis},
  booktitle    = {AAAI-25, Sponsored by the Association for the Advancement of Artificial
                  Intelligence, February 25 - March 4, 2025, Philadelphia, PA, {USA}},
  pages        = {18476--18484},
  publisher    = {{AAAI} Press},
  year         = {2025},
  url          = {https://doi.org/10.1609/aaai.v39i17.34033},
  doi          = {10.1609/AAAI.V39I17.34033},
  bibsource    = {dblp computer science bibliography, https://dblp.org}
}

@inproceedings{murphy2024balor,
  author       = {Emmet Murphy and
                  Lana Josipovic},
  editor       = {Jinjun Xiong and
                  Robert Wille},
  title        = {Balor: {HLS} Source Code Evaluator Based on Custom Graphs and Hierarchical
                  GNNs},
  booktitle    = {Proceedings of the 43rd {IEEE/ACM} International Conference on Computer-Aided
                  Design, {ICCAD} 2024, Newark Liberty International Airport Marriott,
                  NJ, USA, October 27-31, 2024},
  pages        = {227:1--227:9},
  publisher    = {{ACM}},
  year         = {2024},
  url          = {https://doi.org/10.1145/3676536.3676788},
  doi          = {10.1145/3676536.3676788},
  bibsource    = {dblp computer science bibliography, https://dblp.org}
}

@inproceedings{wu2022hls,
  author       = {Nan Wu and
                  Hang Yang and
                  Yuan Xie and
                  Pan Li and
                  Cong Hao},
  editor       = {Rob Oshana},
  title        = {High-level synthesis performance prediction using GNNs: benchmarking,
                  modeling, and advancing},
  booktitle    = {{DAC} '22: 59th {ACM/IEEE} Design Automation Conference, San Francisco,
                  California, USA, July 10 - 14, 2022},
  pages        = {49--54},
  publisher    = {{ACM}},
  year         = {2022},
  url          = {https://doi.org/10.1145/3489517.3530408},
  doi          = {10.1145/3489517.3530408},
  bibsource    = {dblp computer science bibliography, https://dblp.org}
}

@inproceedings{sohrabizadeh2022automated,
  author       = {Atefeh Sohrabizadeh and
                  Yunsheng Bai and
                  Yizhou Sun and
                  Jason Cong},
  editor       = {Michael Adler and
                  Paolo Ienne},
  title        = {Automated Accelerator Optimization Aided by Graph Neural Networks},
  booktitle    = {{FPGA} '22: The 2022 {ACM/SIGDA} International Symposium on Field-Programmable
                  Gate Arrays, Virtual Event, USA, 27 February 2022 - 1 March 2022},
  pages        = {50},
  publisher    = {{ACM}},
  year         = {2022},
  url          = {https://doi.org/10.1145/3490422.3502330},
  doi          = {10.1145/3490422.3502330},
  bibsource    = {dblp computer science bibliography, https://dblp.org}
}

@inproceedings{bai2022improving,
  author       = {Yunsheng Bai and
                  Atefeh Sohrabizadeh and
                  Yizhou Sun and
                  Jason Cong},
  editor       = {Rob Oshana},
  title        = {Improving GNN-based accelerator design automation with meta learning},
  booktitle    = {{DAC} '22: 59th {ACM/IEEE} Design Automation Conference, San Francisco,
                  California, USA, July 10 - 14, 2022},
  pages        = {1347--1350},
  publisher    = {{ACM}},
  year         = {2022},
  url          = {https://doi.org/10.1145/3489517.3530629},
  doi          = {10.1145/3489517.3530629},
  bibsource    = {dblp computer science bibliography, https://dblp.org}
}

@inproceedings{sohrabizadeh2023harp,
  author       = {Atefeh Sohrabizadeh and
                  Yunsheng Bai and
                  Yizhou Sun and
                  Jason Cong},
  title        = {Robust GNN-Based Representation Learning for {HLS}},
  booktitle    = {{IEEE/ACM} International Conference on Computer Aided Design, {ICCAD}
                  2023, San Francisco, CA, USA, October 28 - Nov. 2, 2023},
  pages        = {1--9},
  publisher    = {{IEEE}},
  year         = {2023},
  url          = {https://doi.org/10.1109/ICCAD57390.2023.10323853},
  doi          = {10.1109/ICCAD57390.2023.10323853},
  bibsource    = {dblp computer science bibliography, https://dblp.org}
}

@article{ferretti2022graph,
  title={Graph neural networks for high-level synthesis design space exploration},
  author={Ferretti, Lorenzo and Cini, Andrea and Zacharopoulos, Georgios and Alippi, Cesare and Pozzi, Laura},
  journal={ACM Transactions on Design Automation of Electronic Systems},
  volume={28},
  number={2},
  pages={1--20},
  year={2022},
  publisher={ACM New York, NY}
}

@inproceedings{xu2019powerful,
  author       = {Keyulu Xu and
                  Weihua Hu and
                  Jure Leskovec and
                  Stefanie Jegelka},
  title        = {How Powerful are Graph Neural Networks?},
  booktitle    = {7th International Conference on Learning Representations, {ICLR} 2019,
                  New Orleans, LA, USA, May 6-9, 2019},
  publisher    = {OpenReview.net},
  year         = {2019},
  url          = {https://openreview.net/forum?id=ryGs6iA5Km},
  bibsource    = {dblp computer science bibliography, https://dblp.org}
}

@inproceedings{oono2020graph,
  author       = {Kenta Oono and
                  Taiji Suzuki},
  title        = {Graph Neural Networks Exponentially Lose Expressive Power for Node
                  Classification},
  booktitle    = {8th International Conference on Learning Representations, {ICLR} 2020,
                  Addis Ababa, Ethiopia, April 26-30, 2020},
  publisher    = {OpenReview.net},
  year         = {2020},
  url          = {https://openreview.net/forum?id=S1ldO2EFPr},
  bibsource    = {dblp computer science bibliography, https://dblp.org}
}

@inproceedings{alon2021bottleneck,
  author       = {Uri Alon and
                  Eran Yahav},
  title        = {On the Bottleneck of Graph Neural Networks and its Practical Implications},
  booktitle    = {9th International Conference on Learning Representations, {ICLR} 2021,
                  Virtual Event, Austria, May 3-7, 2021},
  publisher    = {OpenReview.net},
  year         = {2021},
  url          = {https://openreview.net/forum?id=i80OPhOCVH2},
  bibsource    = {dblp computer science bibliography, https://dblp.org}
}

@article{prakriya2025lift,
  author       = {Neha Prakriya and
                  Zijian Ding and
                  Yizhou Sun and
                  Jason Cong},
  title        = {{LIFT:} LLM-Based Pragma Insertion for {HLS} via {GNN} Supervised
                  Fine-Tuning},
  journal      = {CoRR},
  volume       = {abs/2504.21187},
  year         = {2025},
  url          = {https://doi.org/10.48550/arXiv.2504.21187},
  doi          = {10.48550/ARXIV.2504.21187},
  eprinttype    = {arXiv},
  eprint       = {2504.21187},
  bibsource    = {dblp computer science bibliography, https://dblp.org}
}

@inproceedings{masouros2024lbr,
  author       = {Dimosthenis Masouros and
                  Aggelos Ferikoglou and
                  Georgios Zervakis and
                  Sotirios Xydis and
                  Dimitrios Soudris},
  editor       = {Vivek De},
  title        = {Late Breaking Results: Language-level QoR modeling for High-Level
                  Synthesis},
  booktitle    = {Proceedings of the 61st {ACM/IEEE} Design Automation Conference, {DAC}
                  2024, San Francisco, CA, USA, June 23-27, 2024},
  pages        = {351:1--351:2},
  publisher    = {{ACM}},
  year         = {2024},
  url          = {https://doi.org/10.1145/3649329.3663500},
  doi          = {10.1145/3649329.3663500},
  bibsource    = {dblp computer science bibliography, https://dblp.org}
}

@inproceedings{qin2024crossmodality,
  author       = {Zongyue Qin and
                  Yunsheng Bai and
                  Atefeh Sohrabizadeh and
                  Zijian Ding and
                  Ziniu Hu and
                  Yizhou Sun and
                  Jason Cong},
  editor       = {Hussam Amrouch and
                  Jiang Hu and
                  Siddharth Garg and
                  Yibo Lin},
  title        = {Cross-Modality Program Representation Learning for Electronic Design
                  Automation with High-Level Synthesis},
  booktitle    = {Proceedings of the 2024 {ACM/IEEE} International Symposium on Machine
                  Learning for CAD, {MLCAD} 2024, Salt Lake City, UT, USA, September
                  9-11, 2024},
  pages        = {14},
  publisher    = {{ACM}},
  year         = {2024},
  url          = {https://doi.org/10.1145/3670474.3685952},
  doi          = {10.1145/3670474.3685952},
  bibsource    = {dblp computer science bibliography, https://dblp.org}
}

@inproceedings{xiong2024hlspilot,
  author       = {Chenwei Xiong and
                  Cheng Liu and
                  Huawei Li and
                  Xiaowei Li},
  editor       = {Jinjun Xiong and
                  Robert Wille},
  title        = {HLSPilot: LLM-based High-Level Synthesis},
  booktitle    = {Proceedings of the 43rd {IEEE/ACM} International Conference on Computer-Aided
                  Design, {ICCAD} 2024, Newark Liberty International Airport Marriott,
                  NJ, USA, October 27-31, 2024},
  pages        = {226:1--226:9},
  publisher    = {{ACM}},
  year         = {2024},
  url          = {https://doi.org/10.1145/3676536.3676781},
  doi          = {10.1145/3676536.3676781},
  bibsource    = {dblp computer science bibliography, https://dblp.org}
}

@inproceedings{hong2024llm,
  title={Llm-aided compilation for tensor accelerators},
  author={Hong, Charles and Bhatia, Sahil and Haan, Altan and Dong, Shengjun Kris and Nikiforov, Dima and Cheung, Alvin and Shao, Yakun Sophia},
  booktitle={2024 IEEE LLM Aided Design Workshop (LAD)},
  pages={1--14},
  year={2024},
  organization={IEEE}
}

@article{swaroopa2025evaluating,
  title={Evaluating Large Language Models for Automatic Register Transfer Logic Generation for Combinational Circuits via High-Level Synthesis},
  author={Swaroopa, Sneha and Mukherjee, Rijoy and Debnath, Anushka and Chakraborty, Rajat Subhra},
  journal={Foundatiosn and Trends in Electronic Design Automation},
  volume={14},
  number={4},
  pages={295--314},
  year={2025},
  publisher={Emerald Publishing Limited}
}

@article{collini2025c2hlsc,
  title={C2hlsc: Leveraging large language models to bridge the software-to-hardware design gap},
  author={Collini, Luca and Garg, Siddharth and Karri, Ramesh},
  journal={ACM Transactions on Design Automation of Electronic Systems},
  volume={30},
  number={6},
  pages={1--24},
  year={2025},
  publisher={ACM New York, NY}
}

@inproceedings{islam2024gemini,
  title={Gemini-the most powerful LLM: Myth or Truth},
  author={Islam, Raisa and Ahmed, Imtiaz},
  booktitle={2024 5th Information Communication Technologies Conference (ICTC)},
  pages={303--308},
  year={2024},
  organization={IEEE}
}

@inproceedings{mahapatra2014machine,
  title={Machine-learning based simulated annealer method for high level synthesis design space exploration},
  author={Mahapatra, Anushree and Schafer, Benjamin Carrion},
  booktitle={Proceedings of the 2014 Electronic System Level Synthesis Conference (ESLsyn)},
  pages={1--6},
  year={2014},
  organization={IEEE}
}

@inproceedings{sengupta2015user,
  title={User power-delay budget driven PSO based design space exploration of optimal k-cycle transient fault secured datapath during high level synthesis},
  author={Sengupta, Anirban and Bhadauria, Saumya},
  booktitle={Sixteenth International Symposium on Quality Electronic Design},
  pages={289--292},
  year={2015},
  organization={IEEE}
}

@inproceedings{ferretti2018lattice,
  title={Lattice-traversing design space exploration for high level synthesis},
  author={Ferretti, Lorenzo and Ansaloni, Giovanni and Pozzi, Laura},
  booktitle={2018 IEEE 36th International Conference on Computer Design (ICCD)},
  pages={210--217},
  year={2018},
  organization={IEEE}
}

@article{ferretti2018cluster,
  title={Cluster-based heuristic for high level synthesis design space exploration},
  author={Ferretti, Lorenzo and Ansaloni, Giovanni and Pozzi, Laura},
  journal={IEEE Transactions on Emerging Topics in Computing},
  volume={9},
  number={1},
  pages={35--43},
  year={2018},
  publisher={IEEE}
}

@article{10.1145/2390191.2390202,
author = {Xydis, Sotirios and Pekmestzi, Kiamal and Soudris, Dimitrios and Economakos, George},
title = {Compiler-in-the-loop exploration during datapath synthesis for higher quality delay-area trade-offs},
year = {2013},
issue_date = {January 2013},
publisher = {Association for Computing Machinery},
address = {New York, NY, USA},
volume = {18},
number = {1},
issn = {1084-4309},
url = {https://doi.org/10.1145/2390191.2390202},
doi = {10.1145/2390191.2390202},
journal = {ACM Trans. Des. Autom. Electron. Syst.},
month = jan,
articleno = {11},
numpages = {35}
}

\end{document}